\documentclass[journal,12pt,draftclsnofoot,onecolumn]{IEEEtran}

\usepackage{cite}
\usepackage[dvips]{graphicx}
\usepackage{multicol}

\usepackage{amsfonts}
\usepackage{amssymb}
\usepackage{amsmath}
\usepackage{floatflt}

\begin{document}
\title{Geometric Polarimetry $-$ Part I: Spinors and Wave States}
\author{David~Bebbington,~\IEEEmembership{}
        Laura~Carrea,~\IEEEmembership{}
        and~Ernst~Krogager,~\IEEEmembership{Member,~IEEE}% <-this % stops a space
\thanks{%Manuscript received January 20, 2002; revised November 18, 2002.
        This work was supported by the Marie Curie Research Training
Network AMPER (Contract number HPRN-CT-2002-00205).}% <-this % stops a space
\thanks{D. Bebbington and L. Carrea are with the Centre for Remote Sensing \& Environmetrics, Department of Computing and Electronic Systems, University of Essex, U.K. (email: david@essex.ac.uk; lcarrea@essex.ac.uk).}
\thanks{E. Krogager is with the Sensor Department of the DDRE, Copenhagen, Denmark (email: krogager@mil.dk)}}

\maketitle

\begin{abstract}
A new approach to polarization algebra is introduced. It exploits
the geometric properties of spinors in order to represent wave
states consistently in arbitrary directions in three dimensional
space. In this first expository paper of an intended series the
basic derivation of the spinorial wave state is seen to be
geometrically related to the electromagnetic field tensor in the
spatio-temporal Fourier domain. Extracting the polarization state
from the electromagnetic field requires the introduction of a new
element, which enters linearly into the defining relation. We call
this element the \emph{phase flag} and it is this that keeps track
of the polarization reference when the coordinate system is
changed and provides a phase origin for both wave components. In
this way we are able to identify the sphere of three dimensional
unit wave vectors with the Poincar\'{e} sphere.
\end{abstract}


\IEEEpeerreviewmaketitle

\section{Introduction}
%Introduction

\PARstart{T}{he} development of applications in radar polarimetry
has been vigorous over the past decade, and is anticipated to
continue with ambitious new spaceborne remote sensing missions
(for example TerraSAR-X \cite{terrasar} and TanDEM-X
\cite{tandemx}). As technical capabilities increase, and new
application areas open up, innovations in data analysis often
result. Because polarization data are spatial and vectorial as
well as complex, polarimetry possesses a rich mathematical
background that includes not only standard electromagnetics but
also linear algebra and the algebras of Hermitian forms and Lie
groups \cite{cloude:liegroups}.

Whilst the aforementioned areas are well established
mathematically, the development of the formal theory of
polarimetry historically has been somewhat unusual, in the sense
that it has seemed to require some uncommon concepts, such as the
pseudo-eigenvectors of Huynen \cite{huynenPHD}, and more recently
consimilarity transformations \cite{hornjohnson}, \cite{luneburg},
\cite{mottII} have been invoked. There has long been a sense
amongst some researchers that the theory of polarimetry requires
reforming, and a number of efforts to do so have occurred over
time \cite{graves}, \cite{kostinski}, \cite{agrawal},
\cite{hubbert}, \cite{hubbertbringi}, \cite{czyzspie},
\cite{luneburg}, usually evoking controversial responses
\cite{mieras}, \cite{kostinskiR}, \cite{luneburgreply},
\cite{hubbertreply}. What is quite interesting about these
episodes is that expert opinion is often divided over what
precisely is wrong with any new approach.

Clearly any new attempt along this road must be taken not only
with care, but with a view to justifying the need for and
applications of the new approach. In contrast to earlier suggested
changes to the formalism which addressed quite small parts of the
picture (e.g. Graves \cite{graves} and Hubbert
\cite{hubbertbringi}), we have taken a rather radical view. We
believe that it is necessary to start \emph{ab initio} $-$
essentially from the Maxwell equations. We are aware that some
researchers in the discipline see no essential problem, and hence
no need for a reform $-$ and they may feel quite justified in such
an opinion if they continue to obtain results which work in
practice. In response to this type of objection, we would counter
that it is quite possible that in any given application area, a
certain mathematical formulation works without any contradiction
appearing. Problems can occur, however, if one tries to extend a
representation to another application area where the analysis no
longer work; as time goes on, it is a virtual certainty that such
attempts will be made. The underlying causes of problems may be
seen to lie in the very pervasiveness of polarimetry in scientific
applications. The problem lies in the fact that many application
areas in polarimetry either involve special cases, or avoid
certain special cases $-$ for instance, the conventions used in
the optical community for ellipsometry assume a fixed plane for
reference\footnote{Since ellipsometry is a specular optical
technique (the angle of incidence equals the angle of reflection),
the convention is to use as a plane for reference the plane
spanned by the incident and the reflected beams called plane of
incidence \cite{azzam}.}, so that the transition from what radar
polarimetrists would call bistatic to monostatic involve no
ambiguity. In the radar community, on the contrary, a special
convention is normally used such that monostatic scattering in
reciprocal media involves symmetric scattering matrices. But in
moving to the bistatic case, which is becoming an increasingly
topical interest, there may be inequivalent mathematical
representations, not all of which are equally appropriate: we
address this very briefly below.

We are aware that the general polarimetric community is not
primarily interested in mathematical exposition per se, and there
has been some debate as to whether this is the correct forum in
which to introduce our ideas. However, in our view, so much is
wrong with the current state of polarimetric theory that it is
necessary to explain first what is the matter with it, and
preferably to provide an answer. Although this is necessarily
rather theoretical, we strive to keep our development as
approachable as possible. So before moving on to introduce the
substance of our new approach, we offer a selection of concrete
examples that demonstrate that the current conceptual framework in
polarimetry is lacking, and a cause for concern for the rigor of
the discipline. It should also be reasonably clear that many of
the problems are linked, and must relate to something fairly
fundamental. We hope that it may be seen as worthwhile to invest
effort in establishing a workable theoretical basis for
polarimetry.

We note, firstly, a well known identity concerning unitary
matrices, namely,
\begin{equation}\label{UCT}
    \bar{U}^T=U^{-1} \quad \mbox{or}
\end{equation}
\begin{equation}\label{UT}
    U^T=\bar{U}^{-1}, \quad \quad
\end{equation}
where $\bar{U}$ denotes the complex conjugation of the unitary
matrix $U$, and $U^{T}$ the matrix transpose. This relation in
either form appears to have had, and fulfilled, its potential to
cause confusion in the development and application of radar
polarimetry formalism. Let us take three types of transformation
law that occur in application to matrices in the literature:
\begin{equation}\label{sim}
    A\; \rightarrow \; T^{-1}AT
\end{equation}
\begin{equation}\label{cong}
    A\; \rightarrow \; T^{T}AT
\end{equation}
\begin{equation}\label{cons}
    A\; \rightarrow \; \bar{T}^{-1}AT.
\end{equation}
Here, we are not necessarily assuming any particular properties
for transformation matrix $T$. The first of these (\ref{sim}) is
known mathematically as a \emph{similarity}, the second
(\ref{cong}) as a \emph{congruence}, and the third (\ref{cons}) as
a \emph{consimilarity} or conjugate similarity \cite{hornjohnson}.
To understand why each of these might be applied in a given
situation, it is helpful first to see what kind of object gets
mapped to what.

In case (\ref{sim}), if the matrix $A$ operates on a vector
$\mathbf{x}$, geometrically a point in a space, to produce another
vector or point $\mathbf{x}'$
\begin{equation}\label{Ax}
    \mathbf{x}'=A\mathbf{x},
\end{equation}
then the similarity transformation $T$ provides a new
representation $\bar{\mathbf{x}}$ of the point $\mathbf{x}$,
\begin{equation}\label{Tx}
    \mathbf{x}=T\bar{\mathbf{x}}
\end{equation}
such that transformed operations on transformed vectors are
equivalent to operations performed on old vectors. In fact, using
(\ref{Ax}) and (\ref{Tx})
\begin{equation}\label{}
    T\bar{\mathbf{x}}'=AT\bar{\mathbf{x}} \quad \Rightarrow \quad
    \bar{\mathbf{x}}'=T^{-1}AT\bar{\mathbf{x}}=\bar{A}\bar{\mathbf{x}}
\end{equation}
shows that the operation $\bar{A}$ on $\bar{\mathbf{x}}$ follows
the same equation of $A$ on $\mathbf{x}$ if $A$ is transformed
with a similarity transformation. Moreover, the similarity applies
equally correctly to a sequence of transformations $A_1A_2..$.

The geometrical interpretation of a congruence (\ref{cong}) is
rather different. In this case the matrix $A$ describes a
different type of operation altogether that maps a vector
$\mathbf{x}$, geometrically a point, into an object $\mathbf{u}$
known as covector (see \cite{semple}). Geometrically covectors,
according to the dimension of the space being modelled are lines
in two dimensions, planes in three dimensions or hyperplanes in
general $N$ dimensions, defined by $N$ points. In geometry, this
operation is called correlation \cite{semple}:
\begin{equation}\label{}
    \mathbf{u}'=A\mathbf{x}.
\end{equation}
A transformation $T$ which provides a new representation
$\bar{\mathbf{x}}$ of the point $\mathbf{x}$
\begin{equation}\label{Tpoint}
    \mathbf{x}=T\bar{\mathbf{x}}
\end{equation}
acts in a different way when applied to the covector $\mathbf{u}$
(see \cite{semple} for a full description):
\begin{equation}\label{Tplane}
    \mathbf{u}=T^{-1T}\bar{\mathbf{u}}.
\end{equation}
It is very clear, also from the transformation property that
vectors and covectors are completely different objects, belonging
to different spaces. The congruential transformation, obtained
using (\ref{Tpoint}) and (\ref{Tplane})
\begin{equation}\label{}
    T^{-1T}\bar{\mathbf{u}}'=AT\bar{\mathbf{x}}, \quad \Rightarrow
    \quad \bar{\mathbf{u}}'=T^{T}AT\bar{\mathbf{x}}=\bar{A}\bar{\mathbf{x}}
\end{equation}
is different from a similarity because the thing mapped and the
thing mapped to are different types of objects. An important
example of a correlation matrix $A$ is a quadratic form, that
describes a quadratic surface, such as an ellipse in two
dimensions or an ellipsoid in three. A congruence then ensures
that in the transformed space, points lying on the original
geometrical figure will lie on the transformed one. All this
appears far abstracted from polarimetry, but will ultimately
figure prominently in our development of scattering matrices,
which lies out of the scope of the present paper. However, the
form of (\ref{cong}) is instantly recognizable as that for basis
transformation of a backscatter matrix (for example
\cite{kostinski}) $-$ the only problem is to explain why
theoretically such a form may be required. In the standard
formalism the answer is totally obscure. Note, also, that a
sequence of correlations $A_1A_2A_3..$ does not make mathematical
sense, since the domain and range of the transformations are
different types of object. The domain is constituted by proper
vectors and the range by covectors.

The third example (\ref{cons}) which relates to consimilarity is
an example of an antilinear transformation, in which vectors or
points are mapped to a new space of vectors or points that involve
a complex conjugation to correspond. This type of transformation
is well described in the text by Horn and Johnson
\cite{hornjohnson} which is regularly cited in relation to this
type of transformation. It is unfortunate that, as far as we are
aware, there has been no fundamental physical justification for
introducing this into polarimetry. Because of (\ref{UT}), it is
obvious that in the case of a unitary transformation,
consimilarity and congruence are identical. Perhaps because of the
prevalence of Hermitian forms and of unitary transformations in
relation to basis change, the apparatus of quantum mechanics has
had a strong influence over the mathematical development of
polarimetry. But whilst mainstream quantum mechanics fundamentally
involves only unitary operations on Hilbert spaces, polarimetry in
the sense that it is based on classical electromagnetics,
definitely does involve non-unitary processes, because scatterers
and media in general can absorb. We reiterate the point that
consimilarity is not a principle. Also germane to the previous
point is the concept of wave-reversal. It is of course well known
that, when using the same coordinate system to compare two waves
that trace the same polarization ellipse, but propagate in
opposite directions, the rotation of the complex vector reverses,
and that in the analytic signal representation, this can be
expressed by conjugation of the phasor \cite{luneburg}.
Consequently, in the Jones vector representation \cite{jonesI},
the vector components are conjugated for wave reversal. This is
sometimes mistaken as a justification for adopting the
consimilarity formalism in radar backscattering. However, it is
quite clear that backscatter is not an antilinear transformation,
since no time-reversal occurs. Mathematically, the operation of
conjugation cannot be used selectively $-$ a point that can be
missed by the unwary when analytic signal representation is used.
Specifically, for the analytic representation of a harmonic plane
wave, we can write
\begin{equation}\label{}
    \overline{\mbox{e}^{j(\omega t-kz)}}= \overline{\mbox{e}^{j\omega t}}
    \;\overline{\mbox{e}^{-jkz}}=\mbox{e}^{j(kz-\omega t)},
\end{equation}
where the overbar $\overline{\mbox{e}^{j(\omega t-kz)}}$ denotes
the complex conjugate operation, $\omega$ is angular frequency and
$k$ is the wavenumber. The final conjugated result has the same
phase velocity as the unconjugated wave. Mathematical conjugation %$v_p=\frac{\omega}{k}$
cannot be invoked to reverse just space or just time.

One of the most significant and useful developments in
polarimetric analysis has been that of the target vector
covariance theory introduced by Cloude
\cite{cloude:electronicletter}, and which has been developed to a
mature state \cite{cloude:review}. Here, the starting point is a
linear representation of scattering matrices $S$ using the Pauli
matrix decomposition:
\begin{equation}\label{}
   \!\!\!\!\!\!\!\!\!\!\!\!\!\!\!\!\!\!\!\!\!
   S=a\sigma_0+b\sigma_1+c\sigma_2+d\sigma_3=\qquad
   \qquad\qquad\qquad
\end{equation}
\begin{equation} \nonumber
    =a\begin{pmatrix}
      1 & \quad \!\! 0 \\
      0 & \quad \!\! 1 \\
    \end{pmatrix}+b\begin{pmatrix}
      1 & \quad\!0 \\
      0 & -1 \\
    \end{pmatrix}+c\begin{pmatrix}
      0 & \quad \!\! 1 \\
      1 & \quad \!\! 0 \\
    \end{pmatrix}+d\begin{pmatrix}
      0 & -j \\
      j & \quad\!0 \\
    \end{pmatrix},
\end{equation}
where $\sigma_l$ are one matrix representation of the Pauli spin
operators. In this, the representation of the scattering matrix is
taken as a given, and the group-theoretic techniques that have
arisen in this area arguably have more to do with information
theoretic principles than the physics of scattering. If, however,
one wishes to connect this theory to the underlying physical
principles, problems can arise. It was noted in
\cite{bebbington:eusar} that the vectorial interpretation of
scattering matrices can be better related to the Poincar\'{e}
sphere if the original Pauli matrix decomposition is modified.
This problem can be related to the distinction between (\ref{sim})
and (\ref{cong}). Mathematically, Pauli matrices $\sigma_l$ are
defined via multiplicative properties
\begin{equation}\label{sigma_lsigma_m}
    \sigma_l\sigma_m=\delta_{lm}\sigma_0+j\,\varepsilon_{lmn}\sigma_n,
    \quad (l,m,n=0,1,2,3)
\end{equation}
where $\delta_{lm}$ is the Kronecker delta symbol and
$\varepsilon_{lmn}$ is the antisymmetric Levi Civita symbol
\cite{gravitation}. The relation (\ref{sigma_lsigma_m}) means that
the actual realizations of the matrices are just one example of an
equivalence class that transforms according to a similarity
transformation. Geometrically, applying the congruential
transformation (\ref{cong}) to the Pauli matrices makes no sense,
and yet, under a basis transformation, (\ref{cong}) is precisely
what is required. It is also a main plank of the target covariance
theory that it extends to bistatic scattering by an obvious
generalisation, by including the antisymmetric Pauli matrix
\begin{equation}\label{}
    \begin{pmatrix}
      0 & -j \\
      j & \quad\!0 \\
    \end{pmatrix}.
\end{equation}
Geometrically, however, the distinctively different Pauli matrix
is $\sigma_0$. As \cite{bebbington:eusar} shows, an important
simplification arises if a twisted representation is used $-$ and
in fact this can be easily related to the concept of Huynen's
pseudo eigen-equation \cite{huynenPHD}. Target covariance theory
'works' until one tries to relate it to the geometry of the
Poincar\'{e} sphere. A recent example where a misunderstanding of
the underlying mathematics has resulted in a very confused picture
was in \cite{souyris} which attempted to apply a quaternion
analysis to the bistatic case, but neglected the same point. It
should in fact be obvious that, within the backscatter alignment
(BSA) convention, multiplication of scattering matrices makes no
physical sense (for it cannot be used consistently to describe
multiple scatter). Modelling scattering matrices on a division
ring (the technical designation of Pauli matrices \cite{corson})
makes no sense without a fundamental modification.

To summarize the first part of our introduction, we have shown
with a few examples that the mathematical and conceptual framework
that pervades much of the radar polarimetry literature rests on
very shaky foundations. Even if much of the analysis turns out
correctly, both the physical and mathematical principles on which
it rests have become obscured by a great deal of faulty reasoning
over the last half century. We now commence the establishing of a
new framework, which ultimately should unify the treatment of
coherent and incoherent polarimetry in traditional forms and also
naturally extends to the concepts of target covariance. The
anticipated growth in bistatic applications suggests to us that it
is timely to present such a unified theory, particularly, since it
is capable of handling polarization states for waves propagating
in arbitrary directions in the three dimensional space. We also
hope that, once and for all, the origin of apparent idiosyncracies
of polarimetric algebra will be settled.

Before going on to the main exposition, it is appropriate, given
the novelty in the eyes of the polarimetric community of the
formalism we employ, to say a few words concerning the scope of
our objectives here and later, and the importance of what is
gained. The most important single aspect of this work is
emphasised by the first word in the title, which is
\emph{geometric}. This end is far more important than the means,
which is the use of spinor algebra. If the ideas could have been
presented in some other way, we would gladly have done so,
however, it turns out that spinors are so perfectly adapted for
representing polarimetric objects that we would otherwise have had
to re-invent them. Spinors were introduced by Cartan \cite{cartan}
in a geometric context, and may perhaps be best summed up as doing
much the same for geometry as imaginary numbers do for algebra. It
may be remarked that spinors appear rather similar to Jones
vectors, and the question may well arise, as to whether we are
merely re-inventing them. Although these objects appear virtually
identical in some contexts, there is a problem with Jones vectors
in the sense that they seem to have ambiguous properties. In one
sense they are extensions of Euclidean vectors, although they also
may be subjected to unitary (basis) transformations. In
orthogonality relations, Hermitian products are employed, whilst
in antenna voltage equations a Euclidean inner product is
appropriate. Mathematically, this ambiguity makes it difficult to
assign any unique property, and it follows that the mathematical
nature of the polarimetric model as a whole is hard to pin down
rigorously. What is fundamental in this our first paper is the
assignment of a coherent polarimetric state to a spinor, as a
geometric entity, which turns out to be a line in projective
space. Conjugate wave states are represented by conjugate spinors,
and their intersections define Stokes vectors. Because spinors are
truly geometric entities derived from the electromagnetic field,
we are led to a quite outstanding result, that we are able to
identify the Poincar\'{e} sphere with the sphere of normalised
wave-vectors in the Fourier domain. Thus, no longer are
polarization states (in the form of Stokes vectors) consigned to
some abstract space. Rather, we can construct objects all in the
same space. Given this, we can represent arbitrary geometric
rotations in the same analytical framework as other polarimetric
relations.

The key to achieving this will be the introduction of a hitherto
'hidden' object, an implicit reference spinor, which we name the
\emph{phase flag}. This step should enable physicists to overcome
their natural reticence to describe polarization vectors as
spinors, which are in this algebraic form associated in field
theory with 'half-spin' particles like neutrinos. Physicists'
explanations for the $\mathrm{SU}(2)$ representation of light
polarization (e.g. \cite{penroseRTR}), tend to appear half hearted
and lacking conviction. In essence, when a reference spinor is
taken into account, the extra half-spin required by a photon is
accounted for. The amazing simplicity of this idea seems not to
have surfaced in the physics literature, which probably accounts
for why spinors have to date figured little in polarimetry in any
formal sense. This also explains why Stokes vectors do not
transform relativistically \cite{perrin}, even though they appear
to be rather like four vectors.

Our presentation is rigorously derived from Maxwell theory for
harmonic waves, and by the use of a projective formalism that
extends to the spinor case \cite{ruse} provides a complete and
consistent geometric theory of polarization. The importance of a
geometric interpretation for all the objects in polarimetry can
hardly be overstated. For once one has this, algebra is only an
adjunct; geometric relations remain true whatever basis
transformations are applied.

This paper emerges after having worked out the implications of our
ideas to a wide range of polarimetric concepts. Owing to the space
required to set out the fundamentals from first principles, we
limit ourselves here to the modest objective of introducing the
derivation of the formalism as it applies to polarization states.
We have to defer its application to scattering matrices and other
entities which we hope to address in later papers. We can however
assert that the geometrization of polarization leads to very
simple and elegant statements about scattering matrices and their
properties in a most convincing way. Ultimately, it will be seen
that much of the algebra appears quite similar to that in
conventional polarimetry, but the key point is that there is in
the end a clarity of concept and consistency in algebraic
characterization of objects that does not currently obtain.

\section{Spinors and geometry} \label{spinor and geometry}
% Section 1 Spinors and geometry

\begin{figure}[h]
\centering
\includegraphics[scale=0.4]{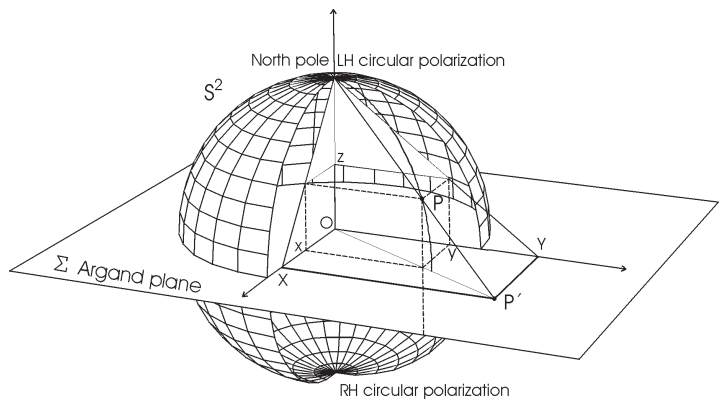}
\caption{The stereographic projection from the Argand plane to the
unit sphere in $\mathbb{R}^3$.} \label{stereo}
\end{figure}

The concept of polarization state has long been associated with
the Poincar\'{e} sphere in a construction sometimes known as a
Hopf mapping through the polarization ratio \cite{mottI}, that is,
fundamentally, a geometric construction. The Hopf map
\cite{nakahara} takes the points $(a,b,c,d)$ on the unit sphere in
$\mathbb{R}^4$, labelled as $S^3=\{(a,b,c,d):a^2+b^2+c^2+d^2=1\}$
to points on the unit sphere in $\mathbb{R}^3$, labelled as
$S^2=\{(x,y,z):x^2+y^2+z^2=1\}$ through the following relations:
\begin{equation}\label{hopf}
      x=2(ac+bd), \quad y=2(bc-ad), \quad z=a^2+b^2-c^2-d^2.
\end{equation}
Introducing $\alpha=a+jb$ and $\beta=c+jd$ we have
\begin{equation}\label{}
    a^2+b^2+c^2+d^2=1 \quad \Rightarrow \quad |\alpha|^2+|\beta|^2=1.
\end{equation}
We can recognize three of the Stokes parameters in $(x,y,z)$ and
the components of the electric field in $(\alpha,\beta)$. Since
\begin{equation}\label{}
    \zeta=X+jY=\frac{x-jy}{1+z}=\frac{c+jd}{a+jb}=\frac{\beta}{\alpha},
\end{equation}
in polarimetry the Poincar\'{e} sphere is recovered through the
stereographic projection, where the polarization ratio,
$\zeta=\frac{\beta}{\alpha}$, is first mapped to a point on the
real Argand plane\footnote{where the real part of $\zeta$ is
represented by a displacement along the $X-$axis and its imaginary
part as a displacement along the $Y-$axis of the Argand plane.}
and then the mapping is normally performed through the
stereographic projection from the Argand plane to the unit
sphere\footnote{ The mapping can be performed in any polarization
representation, since it is a function of a complex variable. In
this case and throughout the paper, we use circular basis,
following the notation of Penrose \cite{penrose}. The polarization
ratio $\zeta$ is chosen as $\zeta=\frac{\beta}{\alpha}$, where the
North pole has a polarization ratio of infinity $(\alpha=0)$
representing left-handed circular polarization and the South pole
has a zero polarization ratio $(\beta=0)$ representing the
right-handed polarization. For a good summary of polarization
ratio in different polarization basis see Mott
\cite{mottI}.\label{nota basi}} as shown in Fig. \ref{stereo}. The
relations (\ref{hopf}) become:
\begin{equation}\label{hopf}
      x=\frac{\zeta+\bar{\zeta}}{\zeta\bar{\zeta}+1}, \quad
      y=\frac{j(\zeta-\bar{\zeta})}{\zeta\bar{\zeta}+1}, \quad z=\frac{1-\zeta\bar{\zeta}}{{\zeta\bar{\zeta}+1}}.
\end{equation}
and they codify the Poincar\'{e} sphere.
%\begin{figure}
%\centering
%\includegraphics[scale=0.4]{stereo}
%\caption{The stereographic projection from the Argand plane to the
%unit sphere in $\mathbb{R}^3$.} \label{stereo}
%\end{figure}

The aim of this paper is to perform a construction of the
Poincar\'{e} sphere as a collection of complex lines and not as a
Hopf map. Each coherent state of polarization will be one of the
complex line generating the sphere, obtained in form of a spinor
from the full electromagnetic field. Such a line can be
parameterized by a complex number equivalent to the polarization
ratio. The first step is now to see how a spinor can be
interpreted as a generating line of a sphere, and to see the links
between a sphere, a complex line and a spinor. We will attach the
physical meaning to the spinor in the next sections to establish
that this sphere is the Poincar\'{e} sphere.

A suitable starting point is the phase of the wave. The analytical
representation of a plane wave is proportional to \cite{jackson}
\begin{equation}\label{}
    \mbox{e}^{j(\omega t-\mathbf{k}\cdot\mathbf{x})}
\end{equation}
with $\omega$ the angular frenquency and
$\mathbf{k}=(k_x,k_y,k_z)$ the wavevector. According to relativity
theory, the phase $\varphi=\omega t-\mathbf{k}\cdot\mathbf{x}$ is
invariant
\cite{jackson}. %pag.529
This means that in two different reference frames (two observers
in uniform relative motion) the plane wave would have different
frequency $\omega$ and wavevector $\mathbf{k}$ but the phase
$\varphi$ would be the same. As a consequence, the invariance of
the phase corresponds to the invariance of a sort of scalar
product between two vectors with four components
$(\frac{\omega}{c},-k_x,-k_y,-k_z)$ and $(ct,x,y,z)$:
\begin{equation}\label{varphi scalare}
    \varphi=\left(\frac{\omega}{c},-k_x,-k_y,-k_z\right)\cdot(ct,x,y,z),
\end{equation}
where $c$ is the speed of light in a vacuum. Because of this
invariance, the frequency and the wavevector of any plane wave
must form a $4-$vector. The tensor description elegantly expresses
the invariance property and highlights the different behavior of
the two $4-$vectors in case of change of reference frame. In fact,
the $4-$vector $(ct,x,y,z)=x^a$ is a true vector since its
dimension is a length, instead the wavevector is a $4-$gradient of
a scalar invariant
\begin{equation}\label{fase onda}
    \left(\frac{\omega}{c},-k_x,-k_y,-k_z\right)=\left(\frac{1}{c}\frac{\partial}{\partial t},\frac{\partial}{\partial x},\frac{\partial}{\partial y},\frac{\partial}{\partial
    z}\right)\varphi=\partial_a \varphi
\end{equation}
which has the dimension of the inverse of a length. The $4-$vector
$x^a$ transforms contravariantly\footnote{which means that the
inverse transformation has to be used for $k_a$.\label{F3}} with
respect to the $4-$vector $k_a=(\frac{\omega}{c},-k_x,-k_y,-k_z)$.
The simplest demonstration of this is that if we change our unit
of length from meters to centimeters, the numerical values of
$x^a$ would scale up while those of $k_a$ would scale down. The
tensor notation automatically takes account of this and
tensors\footnote{A $4-$vector is a type of tensor.} like $x^a$ are
called \emph{contravariant} while tensors like $k_a$ are called
\emph{covariant}. The tensorial notation of the product
(\ref{varphi scalare}) would be:
\begin{equation}\label{}
    \varphi=\sum_{a=0}^{3}k_ax^a\equiv k_ax^a,
\end{equation}
where we have dropped summation sign adopting the Einstein
summation convention where \emph{upper} indices $(x^a)$ are paired
to \emph{lower} $(k_a)$\footnote{Inversely, if the distinction
between contravariant and covariant were not made, scalar products
would only be invariant under isometric transformation such as
rotations or reflections, and not under unitary transformation in
general.}.

We shall try now to obtain representations for the complex lines
generating the sphere. In order to do so, we consider now the
contravariant $4-$vector $x^a=(\tau=ct,x,y,z)$. It is
isomorphic\footnote{An isomorphism (from Greek: isos "equal", and
morphe "shape") is a bijective map between two sets of elements.}
with a Hermitian matrix\footnote{A matrix $\mathrm{M}$ is
Hermitian if $M=M^{\dagger}$ where $ M^{\dagger}$ denotes the
conjugate transpose.} which may be parameterized in the form
\begin{equation}\label{hermitiana}
\begin{pmatrix}
  \tau+z & x+jy \\
  x-jy & \tau-z \\
\end{pmatrix} \qquad \tau,x,y,z \in \mathbb{R}.
\end{equation}
The condition that the matrix be singular may be expressed as
\begin{equation}\label{sfera}
    \tau^2-x^2-y^2-z^2=0.
\end{equation}
In special relativity this condition is satisfied by points lying
on a light-cone through the origin \cite{tonnelat}. An alternative
interpretation is for $(\tau,x,y,z)$ to be considered as
homogeneous projective coordinates \cite{semple}. Let us suppose
we have a point $(X,Y,Z)$ in the Euclidean space $\mathbb{R}^3$.
To represent this same point in the projective space
$\mathbb{P}^3$, we simply add a fourth initial coordinate:
$(1,X,Y,Z)$. These coordinates are the homogenous coordinates of a
point in the projective space $\mathbb{P}^3$. Overall scaling is
unimportant, so the point $(1,X,Y,Z)$ is the same as the point
$(\alpha,\alpha X,\alpha Y,\alpha Z)$\footnote{For any $\alpha\neq
0$, thus the point $(0,0,0,0)$ is disallowed.}. In other words,
\begin{equation}\label{}
    (\tau,x,y,z)\equiv (\alpha \tau,\alpha x,\alpha y,\alpha z).
\end{equation}
Since scaling is unimportant, if the coordinates $(\tau,x,y,z)$
are considered the homogeneous coordinates of a point in the three
dimensional projective space $\mathbb{P}^3$, the equation
(\ref{sfera}) then defines projectively a sphere (or more
generally, a quadric surface):
\begin{equation}\label{sferaunitaria}
    1-\left(\frac{x}{\tau}\right)^2-\left(\frac{y}{\tau}\right)^2-\left(\frac{z}{\tau}\right)^2=0.
\end{equation}
This is reminiscent of the reduction of polarization states of
arbitrary amplitude to a unit Poincar\'{e} sphere.

From one side the vanishing of the determinant of the Hermitian
matrix (\ref{hermitiana}) allows us to build projectively a sphere
but also, since it is Hermitian, to express it \cite{gravitation}
as the Kronecker product
\begin{eqnarray}\label{X}
   %X^{AB^\prime}=
   &&\begin{pmatrix}
  \tau+z & x+jy \\
  x-jy & \tau-z \\
\end{pmatrix}= \\
&=&\begin{pmatrix}
  \xi^0 \\
  \xi^1 \\
\end{pmatrix}\otimes \begin{pmatrix}
  \bar{\xi}^{0^\prime} & \bar{\xi}^{1^\prime}  \\
\end{pmatrix}=\begin{pmatrix}
  \xi^0\bar{\xi}^{0^\prime} & \xi^0\bar{\xi}^{1^\prime} \\
  \xi^1\bar{\xi}^{0^\prime} & \xi^1\bar{\xi}^{1^\prime} \\
\end{pmatrix}  \quad \xi^0,\xi^1 \in \mathbb{C}\nonumber
\end{eqnarray}
where $\otimes$ denotes the Kronecker product, and the overbar
denotes complex conjugation\footnote{The reason for labelling the
conjugated elements with a primed index $\bar{\xi}^{0^\prime}$
will become clear in the sequel section \ref{conjugate}.}. The
matrix on the second line in (\ref{X}) built with the complex
$2-$vector $(\xi^0,\xi^1)$ is in fact singular and Hermitian. The
basic ideas here will appear familiar in polarimetric terms as
relevant to the construction of the coherency matrix \cite{born}.

We can notice that the condition necessary to build a sphere
projectively is the singularity of the matrix and not the
Hermiticity. In fact, if we relax the Hermitian constraint we have
that in general $(\tau,x,y,z)$ are not real, but they still build
a singular matrix and projectively a sphere. The points
$(\tau,x,y,z)$ are complex points lying on the 'real' sphere
defined as
\begin{equation}\label{sphere complex}
    \tau^2-x^2-y^2-z^2=0, \quad \tau,x,y,z \in \mathbb{C}.
\end{equation}
Since the matrix is not Hermitian anymore, we have to redefine the
terms of the Kronecker product (\ref{X})
\begin{eqnarray}\label{Xcomplex}
   %X^{AB^\prime}=
   &&\begin{pmatrix}
  \tau+z & x+jy \\
  x-jy & \tau-z \\
\end{pmatrix}= \\
&=&\begin{pmatrix}
  \xi^0 \\
  \xi^1 \\
\end{pmatrix}\otimes \begin{pmatrix}
  \bar{\eta}^{0^\prime} & \bar{\eta}^{1^\prime}  \\
\end{pmatrix}=\begin{pmatrix}
  \xi^0\bar{\eta}^{0^\prime} & \xi^0\bar{\eta}^{1^\prime} \\
  \xi^1\bar{\eta}^{0^\prime} & \xi^1\bar{\eta}^{1^\prime} \\
\end{pmatrix} \quad \!\!\tau,x,y,z \in \mathbb{C}. \nonumber
\end{eqnarray}
For general complex pairs $(\xi^0,\xi^1)$, $(\eta^0,\eta^1)$, the
matrix is still singular but not Hermitian. Now if
$(\eta^0,\eta^1)$ be fixed, it can be seen that points
\begin{equation}\label{}
    (\tau,x,y,z)
\end{equation}
form a linear one dimensional, complex projective subspace,
$(\xi^0,\xi^1)$ being variable. We want to show now that this one
dimensional projective space is a complex line generating the
sphere. This is the main result of this section. We will use this
conclusion to define the polarization state and the Poincar\'{e}
sphere in the next sections. It should be clear that the same
argument may be applied keeping $(\xi^0,\xi^1)$ constant and
varying $(\eta^0,\eta^1)$. In this way, a second line is
generated. Every single independent $(\xi^0,\xi^1)$
$(\eta^0,\eta^1)$ pair results in a unique point of the sphere.

\begin{figure}[h]
\centering
\includegraphics[scale=0.5]{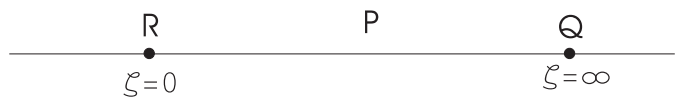}
\caption{A "real" visualization of the complex line. $P=R+\zeta Q$
is any point on the line. For $\xi^1=0$ $(\zeta=0)$, the point $P$
coincides with $R$, and for $\xi^0=0$ $(\zeta=\infty)$ $P\equiv
Q$.} \label{line}
\end{figure}

The singularity of the matrix (\ref{Xcomplex}) implies the
existence of a sphere in the projective space $\mathbb{P}^3$. We
want now to see what is projectively the complex $2-$vector
$(\xi^0,\xi^1)$. As usual, we consider $(\xi^0,\xi^1)$ as
homogeneous coordinate, which means that
$(1,\zeta=\frac{\xi^1}{\xi^0})$ or any non-zero multiples of it,
represents projectively the same vector. So we associate the
vectors $(\xi^0,\xi^1)$ ($\xi^0\neq 0$) with all the complex
numbers together with the infinity $(0,1)$ and we obtain a space
which is usually labelled as $\mathbb{C}\mathbb{P}^1$
\cite{semple}. Regarding the dimension of this space, this depends
on the point of view. From the perspective of the real numbers
already $\mathbb{C}$ is a two-dimensional object often called the
Argand plane\footnote{The space $\mathbb{C}$ differs from
$\mathbb{C}\mathbb{P}^1$ by one point, the infinity.}. As we have
already noticed in polarimetry theory this is the traditional way
to arrive at the Poincar\'{e} sphere, where $\zeta$ might be
interpreted as a polarization ratio. However, there is the other
perspective, the complex perspective, from which $\mathbb{C}$ is
just a one dimensional space. It contains just one complex
parameter. In the same way as $\mathbb{R}$ is the real numbers
line, one would consider $\mathbb{C}$ as the complex numbers line
and so $\mathbb{C}\mathbb{P}^1$ is called the complex projective
line. The elements of $\mathbb{C}\mathbb{P}^1$ are the complex
$2-$vectors $(\xi^0,\xi^1)$, known as \emph{spinors} and
represented by a complex line. The point $P$ on the line is
represented by the projective parameter
$\zeta=\frac{\xi^1}{\xi^0}$ once two reference points $R$ and $Q$
have been chosen (see Fig. \ref{line}):
\begin{equation}\label{}
    P=\xi^0 R+\xi^1 Q, \quad \Rightarrow \quad P=R+\zeta Q.
\end{equation}
%\begin{figure}
%\centering
%\includegraphics[scale=0.5]{line}
%\caption{A "real" visualization of the complex line. $P=R+\zeta Q$
%is any point on the line. For $\xi^1=0$ $(\zeta=0)$, the point $P$
%coincides with $R$, and for $\xi^0=0$ $(\zeta=\infty)$ $P\equiv
%Q$.} \label{line}
%\end{figure}

\begin{figure}[h]
\centering
\includegraphics[scale=0.4]{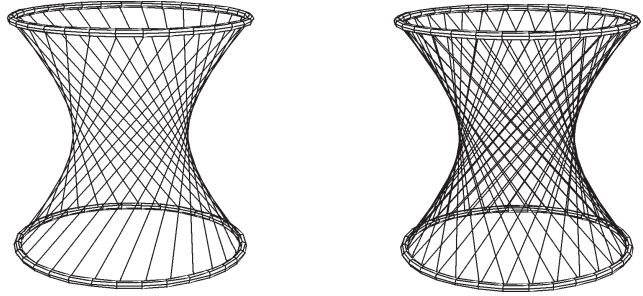}%{hyperboloidI}
\caption{A quadric surface in the form of a hyperboloid. On the
left we can see the hyperboloid generated by one line rotating
around one axis. In the picture on the right we show the two
family of generators.} \label{hyperboloidI}
\end{figure}

At this point we have a complex projective line and a projective
sphere linked by the relation (\ref{Xcomplex}). We want now to
discover the geometric significance of this link: it illustrates
an example in the theory of projective geometry that through any
point on a quadric surface there pass two lines, each of which
lies entirely within the surface \cite{todd}. To those unfamiliar
with complex geometry it may appear surprising that a sphere
\emph{contains} straight lines: this is a fact that escapes us
because on a sphere or any quadric with positive curvature only
one point on each such line is real; all other points are complex.
A simple way to "see" how a complex line can belong to the sphere
is to consider planes intersecting the sphere. We consider for
simplicity a sequence of planes parallel to the $xy-$plane,
cutting the $z-$axis at $z_c$. The equation in the plane of
intersection
\begin{equation}\label{}
    x^2+y^2=1-z_c^2
\end{equation}
is the equation of a circle which degenerates in a point of
tangency ($z_c=\pm1$). In this case, the point of tangency is not
the whole solution. In fact, we have also two intersecting complex
lines with gradient $\pm j$
\begin{equation}\label{}
    x^2+y^2=0 \quad \Rightarrow \quad x=\pm jy
\end{equation}
having one real point $(1,0,0,1)$ at their intersection. What is
surprising is that the two lines lie completely on the sphere
surface. In fact, each point on these lines has coordinate $(1,\pm
jy,y,1)$ which belongs to the sphere. This is a simple example to
illustrate that a sphere contains complex points (\ref{sphere
complex}) and in particular lines built up with complex points!
Now, the next step is to see how such lines can generate the
sphere surface. Since it is difficult to imagine this we can
consider instead a quadric surface with negative
curvature\footnote{In reality, in the complex projective space all
the non-degenerate quadrics are indistinguishable from each other
\cite{semple}.}. For such quadric surfaces, the generating lines
can be wholly real, a fact that is exploited architecturally, e.g.
in the design of cooling towers as cylindrical hyperboloids
\cite{berger}. In Fig. \ref{hyperboloidI} we show how a line can
generate an hyperboloid. We can easily see that i) any point of
the generating line is a point of the surface, ii) there are two
families of generators, and through each point of the surface
there pass two generators, one of each family, iii) generators of
the same family do not intersect, iv) each line of one family
intersects with every other line of the other family.
%\begin{figure}
%\centering
%\includegraphics[scale=0.4]{hyperboloid_two}%{hyperboloidI}
%\caption{A quadric surface in the form of a hyperboloid. On the
%left we can see the hyperboloid generated by one line rotating
%around one axis. In the picture on the right we show the two
%family of generators.} \label{hyperboloidI}
%\end{figure}

The point of the preceding discursive outline is to emphasize the
central principle of geometric polarimetry, which is to identify
the complex $2-$vectors such as
\begin{equation}\label{spinore xi}
   \xi^A=(\xi^0,\xi^1)
\end{equation}
with one of the two sets of complex lines generating the sphere.
The complex $2-$vector $\xi^A$ is known as a
\emph{spinor}\footnote{The term was originated by Cartan
\cite{cartan}.} \cite{vanderwaerden}, \cite{cartan},
\cite{laporte}.  We should stress at this stage that the sphere in
question is to be thought of as the sphere of real unit vectors in
three-dimensional space. The concept of Poincar\'{e} sphere will
be derived later.

In order to establish the notation that we will use in the sequel
of the paper, since the singular matrix (\ref{Xcomplex}) is the
outer product of two spinors $\xi^A$ and the conjugate of
$\eta^B$, we can label it as follows:
\begin{equation}\label{Xconspinore}
    \xi^A\bar{\eta}^{B^\prime}=X^{AB^\prime}\equiv \begin{pmatrix}
  X^{00^\prime} & X^{01^\prime} \\
  X^{10^\prime} & X^{11^\prime} \\
\end{pmatrix}=\begin{pmatrix}
  \tau+z & x+jy \\
  x-jy & \tau-z \\
\end{pmatrix}.
\end{equation}
We can notice that $X^{AB^\prime}$ is considered as a spinor with
two indices; generally spinors can, like tensors have any number
of indices.

We have started from the phase of a plane wave built up with two
$4-$vectors, $x^a$ and $k_a$. But so far we have considered only
the contravariant coordinate $4-$vector $x^a$ in the physical
space. Now since $k_a$ is also a $4-$vector we can think to make
the same considerations we have done for $x^a$. However, there is
a difference. The $4-$vector $k_a$ is covariant. Contravariant and
covariant $4-$vectors are called duals of each other. We want to
explore the meaning of duality in the projective context. First,
we consider the equation of a plane in the three dimensional
Euclidean space:
\begin{equation}\label{piano euclideo}
    f(x,y,z)=ux+vy+wz+q=0.
\end{equation}
The normal vector to the plane is given by
\begin{equation}\label{gradiente piano}
    \mathbf{N}=\mathbf{\nabla}f=(u,v,w).
\end{equation}
The plane may be characterized via its normal vector from the
origin, so a set of three coordinates $(u,v,w)$ can refer to a
plane rather than a point, giving rise to a 'dual' interpretation.
If we consider the projective space, we add a fourth initial
coordinate and we have that the condition:
\begin{equation}\label{piano omo}
    \begin{pmatrix}
  \sigma \\
  u \\
  v \\
  w \\
\end{pmatrix}^T\cdot\begin{pmatrix}
  \tau \\
  x \\
  y \\
  z \\
\end{pmatrix}=\sigma \tau+ux+vy+wz=0
\end{equation}
may be seen as the condition for a plane $(\sigma,u,v,w)$ to pass
through the point $(\tau,x,y,z)$. Comparing (\ref{piano omo}) with
(\ref{piano euclideo}) we have that $q=\sigma \tau$. Since the set
of homogeneous coordinates $(\sigma,u,v,w)$ are defined as a
gradient (\ref{gradiente piano}) we associate it with a covariant
$4-$vector $q_a$ and we associate the set of coordinate
$(\tau,x,y,z)$ interpreted as a point with a contravariant
$4-$vector $x^a$. The linear equation of a plane (\ref{piano omo})
can be written as $q_ax^a=0$, with $q_a=(\sigma,u,v,w)$. Hence the
components of any covariant vector $q_a$ are to be regarded as the
coordinate of a plane.

Reconsidering the phase of the wave (\ref{fase onda}), the zero
phase
\begin{equation}\label{kaxa}
    k_ax^a=0 \quad \Rightarrow \quad \begin{pmatrix}
  \frac{\omega}{c} \\
  -k_x \\
  -k_y \\
  -k_z \\
\end{pmatrix}^T \cdot\begin{pmatrix}
  \tau \\
  x \\
  y \\
  z \\
\end{pmatrix}=0
\end{equation}
in the projective representation  states that the plane $k_a$
passes through the point $x^a$ which belong to the sphere
$1-\left(\frac{x}{\tau}\right)^2-\left(\frac{y}{\tau}\right)^2-\left(\frac{z}{\tau}\right)^2=0$.

At this point, we can make the same consideration we have done for
$x^a$ keeping in mind that $k_a\equiv
(\frac{\omega}{c},-k_x,-k_y,-k_z)$ is a plane.

The singularity of the Hermitian matrix
\begin{equation}\label{KAB}
    K_{AB'}\equiv \begin{pmatrix}
  K_{00^\prime} & K_{01^\prime} \\
  K_{10^\prime} & K_{11^\prime} \\
\end{pmatrix}
\equiv \begin{pmatrix}
  \;\;\;k_0-k_z & -k_x+jk_y \\
  -k_x-jk_y & \;\;\;k_0+k_z \\
\end{pmatrix}.
\end{equation}
defines projectively a sphere
\begin{equation}\label{sfera k}
    1-\left(\frac{k_x}{k_0}\right)^2-\left(\frac{k_y}{k_0}\right)^2-\left(\frac{k_z}{k_0}\right)^2=0
\end{equation}
with $k_0=\frac{\omega}{c}$. The sphere this time is an envelope
of tangent planes $k_a$. The singularity of the matrix (\ref{KAB})
can be expressed as:
\begin{equation}\label{sfera onda}
    k_{a}\Omega^{ab}k_{b}=0 \;
    \mapsto \;
\begin{pmatrix}
  \,k_0 \\
  \,k_1 \\
  \,k_2 \\
  \,k_3 \\
\end{pmatrix}^T
\begin{pmatrix}
  1 & \quad\!0 & \quad\!0 & \quad\!0 \\
  0 & -1 & \quad\!0 & \quad\!0 \\
  0 & \quad \!0 & -1 & \quad\!0 \\
  0 & \quad \!0 & \quad\!0 & -1 \\
\end{pmatrix}
\begin{pmatrix}
  \,k_0 \\
  \,k_1 \\
  \,k_2 \\
  \,k_3 \\
\end{pmatrix}=0
\end{equation}
where $k_1=-k_x$, $k_2=-k_y$, $k_3=-k_z$, so that from here on
numerical as well as symbolic index positions are to be explicitly
interpreted as contravariant $x^a$ or covariant $k_a$ according to
the position. We shall refer to $\Omega^{ab}$ as the wave sphere,
since (\ref{sfera onda}) expresses the Fourier transform of the
wave equation in free space\footnote{As we will see in the section
\ref{maxwell B} in the Fourier domain $\partial_a$ $\rightarrow$
$jk_a$.}. Now we can also give a projective meaning to
$k^a=\Omega^{ab}k_b=(k_0,k^1=k_x,k^2=k_y,k^3=k_z)$ as the point of
tangency of the plane $k_b$ on the sphere $\Omega^{ab}$. In the
Fourier domain, the interpretation of the $4-$vectors $k_a$ as
homogeneous coordinates in three-dimensional projective space
implies normalization of the frequency
\begin{equation}\label{}
    k_a \; \rightarrow \;
    \left(1,\frac{k_1}{k_0},\frac{k_2}{k_0},\frac{k_3}{k_0}\right).
\end{equation}
As we consider harmonic waves individually, and in many
applications a quasi monochromatic assumption is justified this
creates few problems. In return, the interpretation of the linear
algebra as three-, rather then four-dimensional is beneficial from
the point of view of visualization.

As for $X^{AB'}$ in (\ref{Xconspinore}), the singularity of the
matrix (\ref{KAB}), allow us to express it as the Kronecker
product
\begin{equation}\label{KAB con k}
    K_{AB'}=\begin{pmatrix}
     \:K_{00^\prime} & K_{01^\prime} \\
     \:K_{10^\prime} & K_{11^\prime} \\
    \end{pmatrix}=\begin{pmatrix}
      \kappa_0 \\
      \kappa_1 \\
    \end{pmatrix}\otimes \begin{pmatrix}
      \bar{\lambda}_{0^{\prime}} & \bar{\lambda}_{1^{\prime}}\\
    \end{pmatrix}
\end{equation}
this time with the covariant spinors
\begin{equation}\label{}
    \kappa_A=(\kappa_0,\kappa_1) \quad \mbox{and} \quad
    \lambda_A=(\lambda_0,\lambda_1).
\end{equation}
Like the contravariant spinor (\ref{spinore xi}) they also
represent complex lines on the sphere. The concept of duality is
still valid for spinors, namely covariant and contravariant
spinors are still duals of each other\footnote{It means that they
transform covariantly to one another (see footnote \ref{F3}).}.
However, in the three dimensional projective space $\mathbb{P}^3$,
the dual of a line is a line or lines are self-dual. In fact, a
line can be built linking two points but also it is the
intersection of two planes, duals of points. Given two points
$p^a=(p^0,p^1,p^2,p^3)$ and $q^a=(q^0,q^1,q^2,q^3)$ in homogeneous
coordinates, the projective description of the line passing
through the two points are given by the numbers:
\begin{equation}\label{pluecker}
   p^iq^j-p^jq^i
\end{equation}
which build the tensor
\begin{equation}\label{skew line}
    l^{ab}=p^aq^b-q^ap^b.
\end{equation}
 Since $l^{\,ab}=-\,l^{\,ba}$ the tensor is
clearly skew-symmetric:
\begin{equation}\label{skew line matrice}
    l^{\,ab}=\begin{pmatrix}
      0 & -l^{\,10} & -l^{\,20} & -l^{\,30} \\
      l^{\,10} & 0 & l^{\,12} & -l^{\,31} \\
      l^{\,20} & -l^{\,12} & 0 & l^{\,23} \\
      l^{\,30} & l^{\,31} & -l^{\,23} & 0 \\
    \end{pmatrix}
\end{equation}
and therefore the distinct elements are reduced to six
\begin{equation}\label{plueckerCOORD}
    \{l^{\,10},l^{\,20},l^{\,30},l^{\,23},l^{\,31},l^{\,12}\}.
\end{equation}
However, they are not independent since they always satisfy
\begin{equation}\label{linea}
    l^{\,10}l^{\,23}+l^{\,20}l^{\,31}+l^{\,30}l^{\,12}=0
\end{equation}
which is the determinant of a $4\times 4$ matrix
$(p^a,q^a,p^a,q^a)$ that is identically zero. The coordinates
(\ref{plueckerCOORD}) connected by the relation (\ref{linea}) are
called Pluecker (or Grassmann) coordinates of a line. Again
overall scaling is unimportant, namely the set
\begin{equation}\label{}
    \{\alpha\, l^{\,10},\alpha\, l^{\,20},\alpha\, l^{\,30},\alpha\, l^{\,23},\alpha\, l^{\,31},\alpha\, l^{\,12}\}
\end{equation}
represents the same line as (\ref{plueckerCOORD}) does. If the
first coordinate of the points is not zero, it is easy to show
that the coordinates have a nice Euclidean interpretation, namely
\begin{equation}\label{euclidean}
    \begin{array}{l}
      (l^{\,10},l^{\,20},l^{\,30})=\mathbf{p}-\mathbf{q} \\
      (l^{\,23},l^{\,31},l^{\,12})=\mathbf{p}\times \mathbf{q} \\
    \end{array}
\end{equation}
with
$\mathbf{p}=(\frac{p^1}{p^0},\frac{p^2}{p^0},\frac{p^3}{p^0})$,
$\mathbf{q}=(\frac{q^1}{q^0},\frac{q^2}{q^0},\frac{q^3}{q^0})$ and
where $\times$ denotes the cross product. The first set of
coordinates describes the direction of the line from $\mathbf{q}$
to $\mathbf{p}$ and the second describes the plane containing the
line and the origin. The condition (\ref{linea}) is equivalent to
the identically null product
\begin{equation}\label{}
    (\mathbf{p}-\mathbf{q})\cdot (\mathbf{p}\times \mathbf{q})
\end{equation}
where $\cdot$ denotes the scalar product.

Alternatively, we consider the planes $p_a=\Omega_{ab}p^b$ and
$q_a=\Omega_{ab}q^b$. We define the skew-symmetric tensor $l_{ab}$
\begin{equation}\label{skew line dual short}
    l_{ab}=p_aq_b-q_ap_b
\end{equation}
whose components are related to the components of $l^{\,ab}$ in
(\ref{skew line})
\begin{equation}\label{skew line dual}
    l_{ab}=\Omega_{ac}\Omega_{bd}l^{\,cd}=\begin{pmatrix}
      0 & l^{\,10} & l^{\,20} & l^{\,30} \\
      -l^{\,10} & 0 & l^{\,12} & -l^{\,31} \\
      -l^{\,20} & -l^{\,12} & 0 & l^{\,23} \\
      -l^{\,30} & l^{\,31} & -l^{\,23} & 0 \\
    \end{pmatrix}
\end{equation}
which is the dual of the tensor $^{*}r_{ab}$ representing the line
intersection of the planes $p_a$ and $q_a$:
\begin{equation}\label{dual of the line}
    ^{*}r_{ab}=l_{ab}.
\end{equation}
The dual of a tensor is defined through the full antisymmetric
Levi Civita symbol $\varepsilon^{abcd}$ and it is related to the
tensor $r^{\,ab}$ :
\begin{equation}\label{}
    r^{\,ab}=-\frac{1}{2}\,\varepsilon^{abcd}\, ^{*}r_{cd}
\end{equation}
which components can be written in function of the components of
$l^{ab}$:
\begin{equation}\label{}
    r^{\,ab}=\begin{pmatrix}
      0 & -l^{\,23} & -l^{\,31} & -l^{\,12} \\
      l^{\,23} & 0 & -l^{\,30} & l^{\,20} \\
      l^{\,31} & l^{\,30} & 0 & -l^{\,10} \\
      l^{\,12} & -l^{\,20} & l^{\,10} & 0 \\
    \end{pmatrix}.
\end{equation}
The Pluecker coordinates of the line intersection of $p_a$ and
$q_a$ will be the set:
\begin{equation}\label{}
     \{l_{23},l_{31},l_{12},-l_{10},-l_{20},-l_{30}\}.
\end{equation}
Again considering the Euclidean interpretation as in
(\ref{euclidean}), the vector $\mathbf{p}$ and $\mathbf{q}$ now
represents the normal to the planes. For this reason
\begin{equation}\label{euclidean}
    \begin{array}{l}
      (l^{\,23},l^{\,31},l^{\,12})=\mathbf{p}\times \mathbf{q} \\
      (-l^{\,10},-l^{\,20},-l^{\,30})=\mathbf{q}-\mathbf{p} \\
    \end{array}
\end{equation}
this time the first set of coordinates namely the direction of the
line is described by $\mathbf{p}\times \mathbf{q}$ and the second
set namely the plane containing the line and the origin is
described by $\mathbf{q}-\mathbf{p}$. We emphasize that this is
only an Euclidean interpretation that can help us to visualize
things but the Pluecker coordinates are coordinates in the
projective space and not in the Euclidean space.

\section{The algebra of spinors}\label{algebraspinor}
%The algebra of spinors

In the previous section we introduced contravariant and covariant
spinors by interpreting them as representations of the complex
lines generating the sphere. In this section we will discuss the
algebra of spinors especially the relation between covariant and
contravariant spinors and the conjugate spinors.

Spinors were first introduced by Cartan \cite{cartan} in 1913 and
later they were adopted in quantum mechanics to study the
properties of the intrinsic angular momentum of the electron and
other fermions. Today spinors are used in a wide range of branches
of physics and mathematics.

This paper follows the notations and the conventions of the book
by Penrose and Rindler\footnote{which are the same as the
conventions of the book of Misner \cite{gravitation}.}
\cite{penrose} which provides a very clear idea of what spinor
representation signifies.

The spinor $\xi^A$ has four basic forms each of which follows a
separate and distinguished transformation law. This is the
strength of the spinor notation, especially for polarimetry: the
labelling of a spinor of one of the four type immediately dictates
its transformation properties.

Many of the equations of polarimetry involve nothing more
complicated than $2\times 2$ complex matrices. By understanding
the significance of spinor index types, it becomes clear that
these are potentially $4\times 4$ different characters of $2\times
2$ complex matrices; little wonder therefore that confusion may
arise over the correct transformation laws!

We start now to describe the four forms of a spinor and their
transformation law, and later after the derivation of the
polarization state from the electromagnetic field tensor we will
show why and how the distinction of the same object in four forms
is fundamental in polarimetry. The four spinor types can be
categorized as:
\begin{itemize}
    \item the contravariant spinor $\xi^A$,
    \item the covariant spinor $\xi_A$,
    \item the conjugate contravariant spinor $\bar{\xi}^{A'}$,
    \item the conjugate covariant spinor $\bar{\xi}_{A'}$.
\end{itemize}
All these different types have definite and different physical or
geometric significance, and must not be confused with one another.

In the next two sections we describe how to raise and lower
indexes and the conjugation of a spinor. The significance of
raising and lowering is to make a relation between one type and
another. This can be seen as a mapping between objects that are
dual to one another through some fixed geometric property.

\subsection{Index raising and lowering}

In this section we introduce the metric spinor
$\epsilon_{AB}=\epsilon^{AB}$ which tells us how to compute
distances in spinor space. We will see that it is the object
responsible for lowering and raising the indexes and which can
test the independence of two given spinors.

We have seen in the previous section that projectively the
following components of a spinor are equivalent:
\begin{equation}\label{}
    \begin{pmatrix}
      \xi^0 \\
      \xi^1 \\
    \end{pmatrix}\equiv \begin{pmatrix}
      1 \\
      \frac{\xi^1}{\xi^0} \\
    \end{pmatrix}=\begin{pmatrix}
      1 \\
      \zeta \\
    \end{pmatrix}.
\end{equation}
The spinor components $(\xi^0,\xi^1)$ may be interpreted as the
homogeneous coordinate of a point $P$ on a complex line (see Fig.
\ref{line}). Then the complex number $\zeta$ is the inhomogeneous
coordinate of the point on the complex line known as the
\emph{affine} coordinate of a point. Following this terminology we
can see that the difference between complex numbers $\xi$, $\eta$,
namely the affine distance between the two points $P-P'$ may be
expressed in terms of their homogeneous coordinates as
\begin{equation}\label{}
    \xi-\eta=\left[ \begin{pmatrix}
      \quad\!0 & 1 \\
      -1 & 0 \\
    \end{pmatrix}\begin{pmatrix}
      1 \\
      \xi \\
   \end{pmatrix} \right]^T\begin{pmatrix}
      1 \\
      \eta \\
    \end{pmatrix}.
\end{equation}

Now reinterpreting the same relation using the homogeneous
coordinates we can recover the invariant skew-symmetric spinor
$\epsilon_{AB}$ called \emph{metric spinor}:
\begin{equation}\label{}
   \left[ \begin{pmatrix}
      \quad\!0 & 1 \\
      -1 & 0 \\
    \end{pmatrix}\begin{pmatrix}
      \eta^0 \\
      \eta^1 \\
    \end{pmatrix}\right]^T\begin{pmatrix}
      \xi^0 \\
      \xi^1 \\
    \end{pmatrix}\equiv \epsilon_{AB}\xi^A\eta^B,
\end{equation}
where we adopt as usual the Einstein summation
convention\footnote{The expression $\epsilon_{AB}\xi^A\eta^B$
stands for
$\sum_{A=0}^{1}\sum_{B=0}^{1}\epsilon_{AB}\xi^A\eta^B$.}. The
metric spinor has components
\begin{equation}\label{epsilon components}
    \epsilon_{AB} \quad \rightarrow \quad
    \epsilon_{12}=-\epsilon_{21}=1, \quad
    \epsilon_{11}=\epsilon_{22}=0
\end{equation}
namely it is a completely antisymmetric spinor
($\epsilon_{AB}=-\epsilon_{BA}$).

Since the distance is always related to some kind of product, we
can symbolically introduce the inner product
\begin{equation}\label{prodotto interno}
    \epsilon_{AB}\xi^A\eta^B=\xi_B\eta^B,
\end{equation}
where the covariant spinor $\xi_B$
\begin{equation}\label{spinore covariante}
    \xi_B=\epsilon_{AB}\xi^A, \quad \begin{pmatrix}
      \xi_0 \\
      \xi_1 \\
    \end{pmatrix}=\begin{pmatrix}
       -\xi^1 \\
      \quad\!\xi^0 \\
    \end{pmatrix}
\end{equation}
is obtained from the contravariant one with the lowering operator
$\epsilon_{AB}$. Here, as usual, we use the Einstein summation
convention\footnote{which means
$\xi_B=\sum_{A=0}^{1}\epsilon_{AB}\xi^A$, which can be read
$\xi_0=\epsilon_{00}\xi^0+\epsilon_{10}\xi^1$,
$\xi_1=\epsilon_{01}\xi^0+\epsilon_{11}\xi^1$.}.

Likewise, $\epsilon^{AB}$ which has numerically the same element
values as $\epsilon_{AB}$ is a raising operator, performing the
inverse mapping from covariant space to contravariant:
\begin{equation}\label{}
    \xi^A=\epsilon^{AB}\xi_B.
\end{equation}
We note that
\begin{equation}\label{}
    \epsilon_{AB}\,\epsilon^{CB}=\epsilon_{A}^{\;\;C}
\end{equation}
is the Kronecker delta symbol $\delta_{A}^{\;\;C}$ in spinor
space.

It may also be recognized that the product (\ref{prodotto
interno}) vanishes if the ratios of the spinor coordinates of
$\xi=\frac{\xi_1}{\xi_0}$ and $\eta=\frac{\eta_1}{\eta_0}$ are
equal, namely
\begin{equation}\label{prodotto interno zero}
    \xi_B\eta^B=0 \quad \Leftrightarrow \quad \xi^B\propto \eta^B.
\end{equation}
Geometrically, if the two points on the line coincide, the
distance vanishes. So, we have arrived at a symbolic mechanism for
testing the linear independence of two spinors:
\begin{equation}\label{spinori indipendenti}
    \xi_A\eta^A\neq 0.
\end{equation}

We want finally to show how spinors transform for linear
transformations. Since linear transformations map contravariant
spinors to contravariant, we can express the transformation law
for a spinor as
\begin{equation}\label{transf law}
    \xi^A \quad \rightarrow \quad \tilde{\xi}^{A}=L^A_{\;\;
    B}\xi^B.
\end{equation}
with $L^A_{\;\; B}$ the spinor describing the linear
transformation. Using their matrix representation we can write:
\begin{equation}\label{}
    \begin{pmatrix}
      \xi^0 \\
      \xi^1 \\
    \end{pmatrix} \quad \rightarrow \quad \begin{pmatrix}
      \tilde{\xi}^0 \\
      \tilde{\xi}^1 \\
    \end{pmatrix}=\begin{pmatrix}
      L^0_{\;\;
    0} & L^0_{\;\;
    1} \\
      L^1_{\;\;
    0} & L^1_{\;\;
    1} \\
    \end{pmatrix}\begin{pmatrix}
      \xi^0 \\
      \xi^1 \\
    \end{pmatrix},
\end{equation}
where $L^A_{\;\; B}$ is represented by a $2\times 2$ matrix.

\subsection{Conjugate spaces}\label{conjugate}

In the last section we have introduced two types of spinors, the
contravariant and the covariant and their transformation law. The
other two kinds of spinors are the 'primed' ones belonging to the
conjugate space. Thus,
\begin{equation}\label{coniugato}
    \overline{\eta^{A}}=\bar{\eta}^{A^{\prime}}\equiv \begin{pmatrix}
      \bar{\eta}^{0^{\prime}} \\
      \bar{\eta}^{1^{\prime}} \\
    \end{pmatrix}
\end{equation}
in which the numerical values of the components are conjugated,
and the abstract index becomes primed. The conjugate spinor
$\bar{\eta}^{A'}$ will transforms according to the conjugate law
\begin{equation}\label{transf law conj}
    \bar{\tilde{\eta}}^{A'}=\bar{L}^{A'}_{\;\;B'}\bar{\eta}^{B'}
\end{equation}
where $\bar{L}^{A'}_{\;\;B'}$ is the conjugate transformation. The
operation of priming applies equally to covariant spinors, so we
have obtained the four categories of spinors: $\eta^A$, $\eta_A$,
$\bar{\eta}^{A^{\prime}}$, $\bar{\eta}_{A^{\prime}}$.

Using the explicit transformation laws for the four types of
spinors we can infer the transformation law for spinors with more
indices and see why attaching the prime in the index is necessary.

In section \ref{spinor and geometry} we have considered matrices
expressible as a Kronecker product of two spinors
(\ref{Xconspinore})
\begin{equation}\label{X xi eta}
   \xi^A\bar{\eta}^{B^\prime}=X^{AB^\prime}.
\end{equation}

Now using the transformation law (\ref{transf law}) and
(\ref{transf law conj}) for $\xi^A$ and for
$\bar{\eta}^{B^\prime}$ we can show that $X^{AB^\prime}$
transforms not simply as a matrix but
\begin{eqnarray}\label{LXL indici}
    X^{AB'}\quad \rightarrow \quad
    \tilde{X}^{AB'}&=&\tilde{\xi}^A\bar{\tilde{\eta}}^{B^\prime}=L^A_{\;\;C}\,\xi^C\bar{\eta}^{D^\prime}\bar{L}^{B'}_{\;\;D'}=
    \nonumber \\
    &=&L^A_{\;\;C}\,X^{CD'}\,\bar{L}^{B'}_{\;\;D'}.
\end{eqnarray}
Expressing the relation below with the matrix representation we
can rewrite it as:
\begin{equation}\label{}
   X \quad \rightarrow \quad \tilde{X}=LX\bar{L}^{T},
\end{equation}
where $X$ is the matrix representing $X^{AB'}$ and $\bar{L}^{T}$
is the conjugate transpose of the matrix representing
$L^A_{\;\;B}$.

We can easily see that the prime attached to the index is very
essential in order to keep trace of the fact that they transform
with the conjugate transformation. Clearly, since conjugation is
an involutory operation, it means that a double primed label
reverts to the unprimed.

Through an example we can see that if $L$ is a rotation in space,
$X^{AB'}$ (and consequently the $4-$vector) rotates by double the
angle of rotation of a spinor. Since the group of unitary
transformation $\mathrm{SU}(2)$ is the universal covering group of
the orthogonal rotation group $\mathrm{SO}(3)$, this means that
there is a correspondence between elements of $\mathrm{SU}(2)$ and
of $\mathrm{SO}(3)$\footnote{The correspondence $\mathrm{SU}(2) \:
\rightarrow \: \mathrm{SO}(3)$ is a $2:1$ one \cite{penrose}.}.
The unitary transformation can be expressed with the Cayley-Klein
parameters in terms of the Euler angles $\psi$, $\theta$ and
$\varphi$ for the rotation considered \cite{goldstein}:
\begin{equation}\label{U rotazione}
    U^A_{\;\;
    B} \equiv \begin{pmatrix}
      \cos\frac{\theta}{2} \,\mbox{e}^{j\frac{\psi+\varphi}{2}} & j\sin\frac{\theta}{2}\,\mbox{e}^{j\frac{\psi-\varphi}{2}} \\
      j\sin\frac{\theta}{2}\,\mbox{e}^{-j\frac{\psi-\varphi}{2}} & \cos\frac{\theta}{2} \,\mbox{e}^{-j\frac{\psi+\varphi}{2}} \\
    \end{pmatrix}.
\end{equation}

The example is a rotation in space around the $z$ direction by
angle $\alpha$. The unitary spinor is in this case
\begin{equation}\label{}
    U^A_{\;\;
    B} \equiv\begin{pmatrix}
      \mbox{e}^{-j\frac{\alpha}{2}} & 0 \\
      0 & \mbox{e}^{j\frac{\alpha}{2}} \\
    \end{pmatrix}.
\end{equation}
According to (\ref{transf law}), the new spinor $\tilde{\xi}^A$
after the rotation will be:
\begin{equation}\label{rotation about z}
    \tilde{\xi}^A=U^A_{\;\; B}\,\xi^B=\begin{pmatrix}
      \mbox{e}^{-j\frac{\alpha}{2}} \,\xi^0 \\
      \mbox{e}^{j\frac{\alpha}{2}} \,\xi^1\\
    \end{pmatrix}.
\end{equation}
For $\alpha=2\pi$ this rotation brings back to the original value
$X^{AB'}$ (according to (\ref{LXL indici})) but reverses the sign
of the spinor $\xi^A$.

A MATHEMATICA package \cite{bebbington:klein} has been developed
in order to perform symbolic calculation with spinors and tensors
for radar polarimetry using the Penrose-Rindler notation
\cite{penrose}.

In our view, the correct attribution of covariant, contravariant,
and primed or unprimed is crucial to a proper understanding of
polarimetric transformation properties. A particular source of
confusion in polarimetry arises from a restricted symmetry, in
that for any spinor $\xi^A$, $\bar{\xi}_{A^{\prime}}$ has
numerically the same components up to a phase factor as the
'orthogonal' spinor. In fact, two independent spinors satisfy
(\ref{spinori indipendenti}). In particular, if
\begin{equation}\label{}
    \xi_A\eta^A=1,
\end{equation}
the pair $\{\xi^A,\eta^A\}$ forms a spin frame which allows any
spinors to be represented in components\footnote{A spin frame
constitutes what we usually call a basis.}. The spinor $\eta^A$
then will have components\footnote{if we assume that $\xi^A$ has
unit Hermitian norm, namely if
$\xi^0\bar{\xi}^{0^{\prime}}+\xi^1\bar{\xi}^{1^{\prime}}=1$.}
\begin{equation}\label{}
    \eta^A = \begin{pmatrix}
      -\bar{\xi}^{1^{\prime}} \\
      \quad\!\bar{\xi}^{0^{\prime}} \\
    \end{pmatrix}.
\end{equation}
Comparing them with the components of the spinor
$\bar{\xi}_{A^{\prime}}$, obtained using (\ref{spinore
covariante}) and the covariant form of (\ref{coniugato}) we obtain
the components proportional to $\eta^A$:
\begin{equation}\label{}
    \bar{\xi}_{A^{\prime}}=\begin{pmatrix}
      -\bar{\xi}^{1^{\prime}} \\
      \quad\!\bar{\xi}^{0^{\prime}} \\
    \end{pmatrix}.
\end{equation}
This symmetry is equivalent to the property of unitary matrices
that their conjugate transposes are their inverse. This conflation
of properties of different character spinors (which the Jones
vector calculus fails to distinguish) means that if one restricts
consideration to unitary processes the statement of the
transformation rules in the Jones calculus is not uniquely
determined. An example of misinterpretation is that a backward
propagating elliptical state of polarization is considered
conjugated. For example this is implicit in Graves paper
\cite{graves}. Hubbert \cite{hubbertbringi} accepts this as part
of the backscatter alignment convention. Lueneburg \cite{luneburg}
argues its legitimacy by the use of the time reversal symmetry,
when it holds. However, some radar problems involve propagation in
lossy media. If backscatter with propagation through absorptive
media is to be treated within the theory then confusion over this
accidental symmetry must be avoided.

\section{Maxwell's equations and the wave spinor} \label{maxwell}
%Maxwell's equations and the wave spinor.

In the polarimetric literature the state of polarization is
customarily described with reference to the electric field vector,
although in earlier literature it was the magnetic field vector.
Notwithstanding, to divorce one from another in an electromagnetic
wave is artificial since one can never propagate without the
other.

We will start from the electromagnetic field tensor which contains
both the components of the electric and magnetic fields. We will
extract from it the direction of propagation to obtain the
electromagnetic potential. This will be the right quantity to use
to derive the polarization state since we will be able to
establish from it the form of spinor with one index $\psi^A$ that
is usually designated as the coherent polarization state without
recourse to the complefixication of the Euclidean electric field
$\mathbf{E}$. In this way the nature of complex unitary
transformation for rotation in space will become very clear.

In the next section we derive the spinor form of the
electromagnetic field and in the following one we project out from
the electromagnetic potential a spinor containing the polarization
information.

\subsection{Maxwell's equations and the tensor and spinor forms of fields}

Whilst in engineering and applied science, the vector calculus
form of Maxwell's equations is prevalent, they are more concisely
formulated in tensor form. The electromagnetic field tensor
$F_{ab}$ contains the components $(E_x,E_y,E_z)$ of the electric
field $\mathbf{E}$ and $(B_x,B_y,B_z)$ of the magnetic induction
$\mathbf{B}$ \cite{jackson}:
\begin{equation}\label{}
    F_{ab}=\begin{pmatrix}
  \quad\!0 & \quad\!E_x & \quad\!E_y & \quad\!E_z \\
  -E_x & \quad\!0 & -cB_z & \quad\!cB_y \\
  -E_y & \quad\!cB_z & \quad\!0 & -cB_x \\
  -E_z & -cB_y & \quad\!cB_x & \quad\!0 \\
\end{pmatrix}.
\end{equation}
Together with the Hodge dual of $F_{ab}$ \cite{post},
\begin{equation}\label{}
    ^{*}F_{ab}=\begin{pmatrix}
  0 & -cB_x & -cB_y & -cB_z \\
  cB_x & \quad\!0 & -E_z & \quad\!E_y \\
  cB_y & \quad\!E_z & \quad\!0 & -E_x \\
  cB_z & -E_y & \quad\!E_x & \quad\!0 \\
\end{pmatrix},
\end{equation}
and a further field tensor $G_{ab}$ for linear media containing
the components $(D_x,D_y,D_z)$ of the electric displacement
$\mathbf{D}$ and $(H_x,H_y,H_z)$ of the magnetic field
$\mathbf{H}$,
\begin{equation}\label{}
    G_{ab}=\begin{pmatrix}
  \quad\!0 & \quad\!cD_x & \quad\!cD_y & \quad\!cD_z \\
  -cD_x & \quad\!0 & -H_z & \quad\!H_y \\
  -cD_y & \quad\!H_z & \quad\!0 & -H_x \\
  -cD_z & -H_y & \quad\!H_x & \quad\!0 \\
\end{pmatrix},
\end{equation}
the Maxwell equations in S.I. units can be written as:
\begin{eqnarray}
    \partial_a\,^{*}F^{ab}&=&0 \qquad
    \Rightarrow \quad \left\{ \begin{array}{l}
      \nabla\cdot \mathbf{B}=0 \\
      \nabla \times \mathbf{E}=-\frac{\partial \mathbf{B}}{\partial t} \\
    \end{array} \right.\\
    \partial_a\;G^{ab}&=&J^b \;\;\quad
    \Rightarrow \quad \left\{ \begin{array}{l}
      \nabla\cdot \mathbf{D}=\rho \\
      \nabla \times \mathbf{H}=\mathbf{J}+\frac{\partial\mathbf{D}}{\partial t} \\
    \end{array} \right.
\end{eqnarray}
where $J^b=(c\rho,J_x,J_y,J_z)$ is the current 4-vector with
$\rho$ the charge density and $\mathbf{J}$ the current density and
$\partial_a=\frac{\partial}{\partial
x^a}=(\frac{1}{c}\frac{\partial}{\partial
t},\frac{\partial}{\partial x},\frac{\partial}{\partial
y},\frac{\partial}{\partial z})$ as in (\ref{fase onda}). We
continue to use the Einstein summation convention: when an index
appears twice, once in an upper (superscript) and once in a lower
(subscript) position, it implies that we are summing over all of
its possible values. So for example $\partial_a\,^{*}F^{ab}$
implies $\sum_{a=0}^3\partial_a\,^{*}F^{ab}$. It is clear that the
field $F_{ab}$ is a skew symmetric tensor, namely
$F_{ab}=-F_{ba}$, containing six independent real components, the
$\mathbf{E}$ and $\mathbf{B}$ components. We can then interpret it
as a projective line, since the condition (\ref{linea}) is
satisfied by plane wave radiating fields. In fact, this condition
can be expressed as $^{*}F^{ab}F_{ab}=0$ and it corresponds to
$\mathbf{E}\cdot c\mathbf{B}=0$. Comparing the tensor $F^{ab}$
\begin{equation}\label{}
    F^{ab}=\begin{pmatrix}
  \quad\!0 & -E_x & -E_y & -E_z \\
  E_x & \quad\!0 & -cB_z & \quad\!cB_y \\
  E_y & \quad\!cB_z & \quad\!0 & -cB_x \\
  E_z & -cB_y & \quad\!cB_x & \quad\!0 \\
\end{pmatrix}.
\end{equation}
with (\ref{skew line}) we can write the Pluecker coordinates of
the line (\ref{plueckerCOORD}):
\begin{equation}\label{}
    (E_x,E_y,E_z,-cB_x,-cB_y,-cB_z)
\end{equation}
which are the $\mathbf{E}$ and $\mathbf{B}$ cartesian components.

In spinor form, a real electromagnetic field may be represented by
a mixed spinor \cite{penrose},
\begin{equation}\label{FABAB}
    F_{ABA^{\prime}B^{\prime}}=\varphi_{AB}\,\epsilon_{A^{\prime}B^{\prime}}+\epsilon_{AB}\,\bar{\varphi}_{A^{\prime}B^{\prime}}.
\end{equation}
where $\epsilon_{AB}$ is the spinor metric defined in
(\ref{epsilon components}) and $\varphi_{AB}$ is called the
electromagnetic spinor. Because $\epsilon_{A^{\prime}B^{\prime}}$
and $\epsilon_{AB}$ are constant spinors and
$\bar{\varphi}_{A^{\prime}B^{\prime}}$ is the conjugate of
$\varphi_{AB}$, this explains why this $\varphi_{AB}$ as symmetric
spinor encodes all the information. Since
$\varphi_{AB}=\varphi_{BA}$ is symmetric, it has three independent
complex components which are related to \cite{penrose}
\begin{equation}\label{}
    \mathbf{E}-jc\mathbf{B}
\end{equation}
for real field-vectors $\mathbf{E}$, $\mathbf{B}$. In matricial
form, $\varphi_{AB}$, a spinor with two indices like
$\epsilon_{AB}$ is a two-by-two matrix.

The information of the field is bundled in a two-index entity, the
electromagnetic spinor $\varphi_{AB}$. From the quantum physical
standpoint, the electromagnetic field is carried by photons,
particles of spin equal to one. If we want to represent a quantum
field with a spinor, a simple rule for the number of spinor
indices is that the number of indices of like type is twice the
quantum spin. So $\varphi_{AB}$ automatically represents a spin
$1$ boson such as a photon. It is also the case that a spinor with
primed indices $\bar{\varphi}_{A^{\prime}B^{\prime}}$ represents
the antiparticle field, and in case of a photon the opposite
helicity.

Our aim is to attach a physical meaning to the spinor with one
index of section \ref{spinor and geometry} that is to derive the
polarization state in the form of spinor with one index $\psi^A$
that is usually designated as the coherent polarization state.
However, a spinor with one index, like $\psi^A$ would represent
quantum fields for fermions with spin equal to $\frac{1}{2}$, e.g.
the massless neutrino. Therefore from the physicist's point of
view it may appear manifestly incorrect to represent an
electromagnetic field by a spinor with one index\footnote{except
in a two dimensional representation \cite{berry96}.}.

The solution to this problem, which explains why the polarimetric
notation of transverse wave states and Stokes vectors has not to
our knowledge been formalized in spinor form, is that there is
missing structure. Absence of this missing structure from the
representation means that polarimetric tensor and spinor
expressions do not appear to transform correctly. From the
polarimetrist's point of view, absence of the missing structure
leaves room for ambiguity in the way Jones vectors should be
transformed when the direction of propagation is variable,
something already highlighted in the work of Ludwig \cite{ludwig}.

In order to obtain the polarization state in one index spinor and
see where the missing structure is hidden we need to consider the
electromagnetic potential as the primary object rather than the
field tensor.

\subsection{The electromagnetic potential and the wave
spinor}\label{maxwell B}

The expression (\ref{FABAB}) is not suitable to derive a one-index
spinor to represent a harmonic polarization state since the
polarization information is equally contained in $\varphi_{AB}$
and $\bar{\varphi}_{A^{\prime}B^{\prime}}$. For this
representation, a more convenient spinor to use, which carries all
the required information is the electromagnetic
potential\footnote{The vector potential is often used in antenna
theory, where the potential in the far field can be simply related
to each current element in the source.}. It is a $4-$vector whose
components are the electrostatic potential $\phi$ and the magnetic
vector potential $\mathbf{A}$:
\begin{equation}\label{}
    \Phi^a\equiv (\phi, \mathbf{A}).
\end{equation}

The electromagnetic tensor $F_{ab}$ can be expressed in terms of
$\Phi_a$ as \cite{jackson}:
\begin{equation}\label{F con potenziale}
    F_{ab}=\partial_a \Phi_b-\partial_b\Phi_a \quad \Rightarrow
    \quad \left\{\begin{array}{l}
      \mathbf{E}=-\frac{1}{c}\frac{\partial \mathbf{A}}{\partial t}- \nabla \phi \\
      \mathbf{B}=\nabla \times \mathbf{A} \\
    \end{array} \right.
\end{equation}
where $\Phi_a\equiv (\phi,-\mathbf{A})$. If we restrict attention
to the Fourier domain then the derivative becomes simply an
algebraic operation. In fact,
\begin{equation}\label{fourier domain}
    \partial_a \; \rightarrow \; jk_a
\end{equation}
and the (\ref{F con potenziale}) becomes the skew symmetrized
outer product,
\begin{equation}\label{F}
    F_{ab}=j(k_a\Phi_b-k_b\Phi_a).
\end{equation}
Note that, from now on, since we are working in the Fourier
domain, the harmonic analytical signal representation is implicit.
The fields and the derived objects become complex.

At this point we can notice that, given the potential, in order to
derive the field, $F_{ab}$, it is necessary to assume the wave
vector, $k_a$. However, in polarimetry, this is the goal: to strip
off assumed or known quantities: frequency, direction of
propagation, even amplitude and phase, to arrive at the
\emph{polarization state}. Now, in spinor form, the vector
potential $\Phi_a$ is a $4-$vector isomorphic to an hermitian
matrix, $\Phi_{AB^{\prime}}$ like any covariant $4-$vector
(\ref{KAB})
\begin{equation}\label{pot spinore}
    \Phi_{AB^{\prime}}\equiv \begin{pmatrix}
  \;\;\;\phi-A_z & -A_x+jA_y \\
  -A_x-jA_y & \;\;\;\phi+A_z \\
\end{pmatrix}.
\end{equation}
Moreover, as is well known, the vector potential has gauge
freedom, so is not uniquely specified for a given field. In fact,
the field $F_{ab}$\footnote{and consequently $\mathbf{E}$ and
$\mathbf{B}$.} is not altered if the potential $\Phi_{a}$ is
changed subtracting the $4-$gradient of some arbitrary function
$\chi$:
\begin{equation}\label{}
   \tilde{\Phi}_a=\Phi_a-\partial_a \chi \quad \Rightarrow \quad \left\{\begin{array}{l}
     \tilde{\phi}= \phi-\frac{\partial \chi}{\partial t} \\
     \tilde{\mathbf{A}}= \mathbf{A}+\nabla \chi \\
   \end{array} \right..
\end{equation}

In relativity texts, a particular gauge, namely the Lorenz gauge,
is typically singled out, as it is a unique choice that is
invariant under Lorentz transformations. However, if this
generality is not required other choices are possible:
\begin{itemize}
    \item the radiation gauge, also known as the transverse gauge,
    has the consequence that
    \begin{equation}\label{rad}
    k^a F_{ab}=0,
    \end{equation}
    \item the Coulomb gauge can be expressed as
    \begin{equation}\label{coulomb}
    \omega_a\Phi^a=0,
    \end{equation}
    with $\omega_a\equiv (1,0,0,0)$, and implies $\phi=0$.
\end{itemize}
These two choices are not mutually exclusive and can be
simultaneously satisfied for a radiating plane wave. Then we have
for a wave propagating in the $z-$direction that the radiation
gauge condition implies that $A_z=0$ and together with the Coulomb
gauge we obtain
\begin{eqnarray}\label{phia}
   \Phi_a&\equiv& (0,-A_x,-A_y,0) \;\rightarrow \; \Phi_{AB^{\prime}}=\begin{pmatrix}
      0 & \Phi_{01^{\prime}} \\
      \Phi_{10^{\prime}} & 0 \\
    \end{pmatrix}= \nonumber \\
   &=&\begin{pmatrix}
      0 & -A_x+jA_y \\
      -A_x-jA_y & 0 \\
    \end{pmatrix}.
\end{eqnarray}
This contains two generally non-vanishing components that
transform conjugately with respect to one another, and can be
identified with components of opposite helicity (the two circular
polarization components), since the $\mathbf{E}$ vector is
algebraically proportional to the vector potential in the Fourier
domain (\ref{F con potenziale}). In order to obtain the
traditional 'Jones vector' representation as the one-index spinor
$\psi_A$, it is necessary to find a form of projection of
$\Phi_{AB^{\prime}}$ onto a spinor with one index. The
polarization information contained in (\ref{phia}) can be
amalgamated into a single spinor simply by contracting with a
\emph{constant} spinor $\bar{\theta}^{B^{\prime}}$:
\begin{equation}\label{phitheta}
    \psi_A=jk_0 \,\Phi_{AB^{\prime}}\bar{\theta}^{B^{\prime}}=jk_0\,\begin{pmatrix}
      0 & \Phi_{01^{\prime}} \\
      \Phi_{10^{\prime}} & 0 \\
    \end{pmatrix}\begin{pmatrix}
      1 \\
      1 \\
    \end{pmatrix}=\begin{pmatrix}
      jk_0\,\Phi_{01^{\prime}} \\
      jk_0\,\Phi_{10^{\prime}} \\
    \end{pmatrix},
\end{equation}
with $k_0=\frac{\omega}{c}\,$. This achieves the stated goal and
$jk_0\,\Phi_{01^{\prime}}$ and $jk_0\,\Phi_{10^{\prime}}$ are the
circular polarization components\footnote{In fact in the Fourier
representation, the $\mathbf{E}$ vector is algebraically
proportional to the vector potential $\mathbf{A}$. Using (\ref{F
con potenziale}) and (\ref{fourier domain}) we obtain
$\mathbf{E}=-jk_0\mathbf{A}$. Then,
$jk_0\,\Phi_{01^{\prime}}=E_x-jE_y$ and
$jk_0\,\Phi_{10^{\prime}}=E_x+jE_y$ are the left and right
circular polarization components.}. It has been necessary to
introduce extra structure, namely the contraction with
$\bar{\theta}^{B^{\prime}}$ to obtain the one-index representation
$\psi_A$. This explains the apparent physical inappropriateness of
the one-index spinor representation. While the extra structure may
for obvious reasons be unattractive to the theoretical physicists,
it is by contrast of value to the practical polarimetrists because
it can be inserted when needed to resolve questions relating to
the geometry of scattering.

We have to notice that the choice of the constant spinor
$\bar{\theta}^{B^{\prime}}$ in (\ref{phitheta}) has a degree of
arbitrariness. In fact a rotation by an angle $\alpha$ about the
$z-$axis is given by (\ref{rotation about z}):
\begin{equation}\label{theta ruotato}
    \tilde{\bar{\theta}}^{A'}=\bar{U}^{A'}_{\;\;
    B'}\bar{\theta}^{B'}=\begin{pmatrix}
      \mbox{e}^{j\frac{\alpha}{2}}\,\bar{\theta}^{0'} \\
      \mbox{e}^{-j\frac{\alpha}{2}}\,\bar{\theta}^{1'} \\
    \end{pmatrix}.
\end{equation}
Using the new $\tilde{\bar{\theta}}^{B'}$ in the defining relation
(\ref{phitheta}), the new spinor will be
\begin{equation}\label{psi nuova}
    \tilde{\psi}_A=jk_0\,\begin{pmatrix}
      \mbox{e}^{-j\frac{\alpha}{2}}\,\bar{\theta}^{1'}\Phi_{01^{\prime}} \\
      \mbox{e}^{j\frac{\alpha}{2}}\,\bar{\theta}^{0'}\Phi_{10^{\prime}} \\
    \end{pmatrix},
\end{equation}
which shows that a rotation around the $z-$axis will result in a
differential phase offset between the two circular polarization
components $\Phi_{01^{\prime}}$ and $\Phi_{10^{\prime}}$. However,
if unitary property are to be preserved then
\begin{equation}\label{}
    |\tilde{\psi}_0|^2+|\tilde{\psi}_1|^2=|\psi_0|^2+|\psi_1|^2
\end{equation}
which projectively means that the components of
$\bar{\theta}^{B^{\prime}}$ have to be equal in amplitude:
\begin{equation}\label{condizione su theta}
    |\bar{\theta}^{0^{\prime}}|=|\bar{\theta}^{1^{\prime}}|.
\end{equation}
Spinors with equal amplitude components necessarily correspond to
real points
\begin{equation}\label{}
    \theta^b=\theta^B\bar{\theta}^{B^{\prime}}
\end{equation}
on the equatorial plane\footnote{Note that this fact is valid in
circular basis.}. For reasons that will shortly become clear, we
call $\bar{\theta}^{B^{\prime}}$ the \emph{phase flag}.

Whilst (\ref{phia}) employs a special choice of coordinates, it is
important to note that a rotation of reference frame may be
effected through the relation (see (\ref{LXL indici}))
\begin{equation}\label{}
    \tilde{\Phi}_{AB'}=U_A^{\;\;C}\bar{U}_{B'}^{\;\;D'}\Phi_{CD'}.
\end{equation}
The covariant (coordinate independent) formulation of
(\ref{phitheta}) means that it is valid for waves propagating in
arbitrary directions, namely it is valid in any rotated frame if
all the elements are transformed according to the appropriate
rule:
\begin{equation}\label{psi tilde}
    \tilde{\psi}_A=U_A^{\;\;B}\psi_B,
\end{equation}
\begin{equation}\label{}
    \tilde{\Phi}_{AB^{\prime}}=U_A^{\;\;C}\bar{U}_{B^{\prime}}^{\;\;D^{\prime}}\Phi_{CD^{\prime}}
\end{equation}
\begin{equation}\label{teta}
    \tilde{\bar{\theta}}_{A^{\prime}}=\bar{U}_{A^{\prime}}^{\;\;B^{\prime}}\bar{\theta}_{B^{\prime}}
\end{equation}
with $U_A^{\;\;B}$ the unitary spinor describing the rotation of
the frame which components can be parametrized as in (\ref{U
rotazione}).

This is the most important result of this section: we have been
able to define the spinor containing the polarization information
from the electromagnetic field using a constant spinor we named
phase flag. This definition is valid in any reference frame,
namely it is valid for any direction of propagation considered.
Table \ref{table} summaries the main steps to follow to extract
the polarization information from the electromagnetic tensor.

\section{Spinors, phase and the Poincar\'{e} sphere}\label{phasepoincare}
%Spinors, phase and Poincar\'{e} sphere

\begin{figure}[h]
\centering
\includegraphics[scale=0.35]{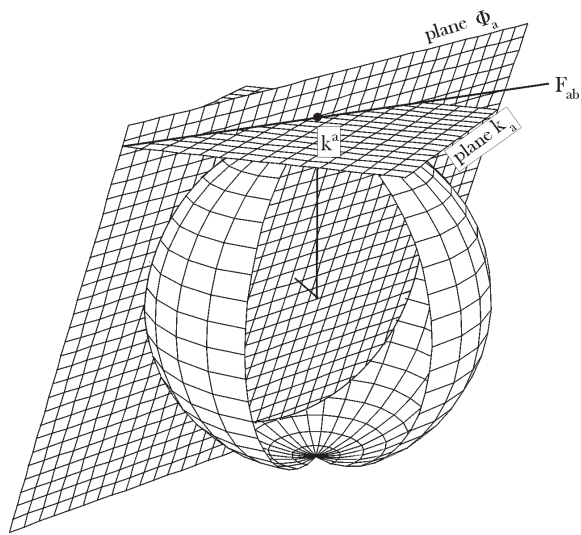}
 \caption{Relation between wave vector, electromagnetic field and
 the potential for the radiation gauge. In this gauge,
 the plane $\Phi_a$ is any plane intersecting the point $k^a$
 on the wave sphere $\Omega^{ab}$, hinging around the line $F_{ab}$ which is the intersection between the plane $\Phi_a$ and the wave plane $k_a$. }
\label{img01}
\end{figure}

\begin{figure}[h]
\centering
\includegraphics[scale=0.35]{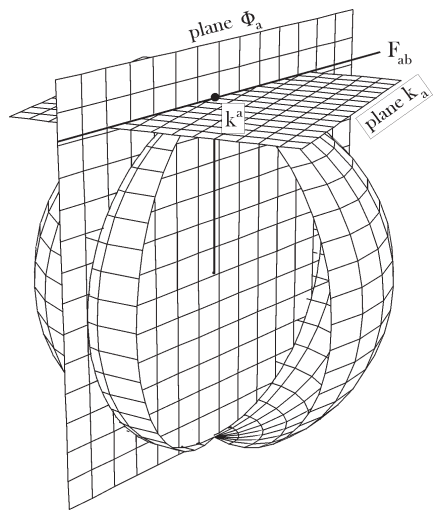}
 \caption{Relation between wave vector, electromagnetic field and the potential
 for the Coulomb gauge. In this gauge, $\Phi_a$ is the plane through the point $k^a$ and the center of the wave sphere $\Omega^{ab}$.
 The electromagnetic field is the same line of Fig. \ref{img01}, the intersection
 between the plane $\Phi_a$ and the wave plane $k_a$. }
\label{img02}
\end{figure}

So far we have deliberately avoided more than scant reference to
the Poincar\'{e} sphere. The reason for this is that by
integrating the material from the previous section with the
projective interpretation of Sections \ref{spinor and geometry}
and \ref{algebraspinor} we are now able to arrive at the first
remarkable consequence of this work, namely that the Poincar\'{e}
sphere may be \emph{identified} with the wave sphere by a direct
geometrical construction. In order to see this, we have simply to
interpret geometrically the algebraic relations of the previous
section. The vector potential in its covariant form $\Phi_a$ may
be represented projectively as a plane as we have seen in the
section \ref{spinor and geometry}. If no gauge condition is
specified, then $\Phi_{a}$ would represent a general plane in
projective space. However, the radiation gauge condition
$k^aF_{ab}=0$ (\ref{rad}) implies that
\begin{equation}\label{}
    k^a\Phi_a=0 \quad \Rightarrow \quad
    \frac{\omega}{c}\phi-k_xA_x-k_yA_y-k_zA_z=0
\end{equation}
namely that $\Phi_a$ is any plane that passes through the point of
tangency of the wave plane $k_a$ with the wave sphere
$\Omega^{ab}$ as shown in Fig. \ref{img01} (see (\ref{kaxa}) and
following).
%\begin{figure}
%\centering
%\includegraphics[scale=0.35]{img01}
% \caption{Relation between wave vector, electromagnetic field and
% the potential for the radiation gauge. In this gauge,
% the plane $\Phi_a$ is any plane intersecting the point $k^a$
% on the wave sphere $\Omega^{ab}$, hinging around the line $F_{ab}$ which is the intersection between the plane $\Phi_a$ and the wave plane $k_a$. }
%\label{img01}
%\end{figure}
The Coulomb gauge condition $\omega_a\Phi^a=0$ (\ref{coulomb}),
however, imposes the condition that $\Phi_a$ passes through the
center of the wave sphere $\omega_a\equiv (1,0,0,0)$. Taken
together, these conditions imply that the plane $\Phi_a$ passes
through the axis of the sphere normal to the wave plane $k_a$
(Fig. \ref{img02}).
%\begin{figure}
%\centering
%\includegraphics[scale=0.35]{img02}
% \caption{Relation between wave vector, electromagnetic field and the potential
% for the Coulomb gauge. In this gauge, $\Phi_a$ is the plane through the point $k^a$ and the center of the wave sphere $\Omega^{ab}$.
% The electromagnetic field is the same line of Fig. \ref{img01}, the intersection
% between the plane $\Phi_a$ and the wave plane $k_a$. }
%\label{img02}
%\end{figure}
The appropriateness of the projective interpretation is seen by
noting that the significance of the line intersection of the
planes $\Phi_a$ and $k_a$ is that it represents projectively the
electromagnetic field tensor $F_{ab}$.

Let us suppose that the field tensor $F_{ab}$ is given.
Algebraically, using homogeneous coordinates, $F_{ab}$ (\ref{F})
in the Fourier domain is proportional to
\begin{equation}\label{Fp}
    F_{ab} \propto k_a \Phi_b-k_b \Phi_a
\end{equation}
and it expresses the dual homogeneous coordinates of a line
(\ref{skew line dual short})-(\ref{dual of the line}), called
Pluecker coordinates\footnote{In general $F_{ab}$ is a bivector
\cite{boulanger} because of its antisymmetry property. For a plane
wave radiating field the bivector becomes simple and it is
representable as the line (\ref{Fp}). The condition for the
bivector $F_{ab}$ to be simple is $\tilde{F}_{ab}F^{ab}=0$ which
is equivalent to $\mathbf{E}\cdot c\mathbf{B}=0$ and to
(\ref{linea}). The condition for the bivector to be null is
$F_{ab}F^{ab}=0$ which is equivalent to
$|\mathbf{E}|^2=|c\mathbf{B}|^2$. This condition states that the
bivector intersects the sphere $\Omega^{ab}$. Both conditions are
met by a plane wave radiating field.} in the projective space
$\mathbb{P}^3$ \cite{semple}.

Now, given the line $F_{ab}$ and the wavevector $k_a$ we have that
$\Phi_a$ is a general plane in $\mathbb{P}^3$. Imposing the
radiation gauge condition (\ref{rad}) we have that the plane
$\Phi_a$ passes through the point $k^a$ but it has still the
freedom to pivot about the line $F_{ab}$ (see Fig. \ref{img01}).
The unique representation is then obtained by choosing  the one
element of the family of planes $\Phi_a$ that passes through the
center of the sphere, namely the plane which satisfy the Coulomb
condition (\ref{coulomb}). Very nicely the geometry neatly
illustrates the gauge freedom.

We now come to our main objective: to express the one index spinor
$\psi_A$ geometrically. In the geometric interpretation, the
totality of wave states with wave vector $k^a$ are obtained by the
lines $F_{ab}$ that pass through the $k^a$ axis in the plane
$k_a$. Note that we may admit complex lines as the geometry is
generally valid over complex coordinates. For every line $F_{ab}$,
the plane $\Phi_a$ is thus uniquely determined given the gauge
fixing.

Now we consider the spinor equivalent of the plane $\Phi_a$ which
is $\Phi_{AB'}$ as in (\ref{pot spinore}) still geometrically
represented as a plane. As seen in the section \ref{spinor and
geometry}, spinors can be interpreted as complex lines generating
the wave sphere (see equations (\ref{KAB})-(\ref{KAB con k})). We
suppose that one of the two generators of the wave sphere
$\bar{\theta}^{B^{\prime}}$ is chosen as a fixed reference
generator. Then provided $\bar{\theta}^{B^{\prime}}$ does not lie
in the plane $\Phi_{AB^{\prime}}$ (which is guaranteed by the
gauge choice), there is a unique point of the generator
$\bar{\theta}^{B^{\prime}}$ that intersects $\Phi_{AB^{\prime}}$
for any wave state and thus we can write
\begin{equation}\label{psi}
    \psi_A=\Phi_{AB^{\prime}}\bar{\theta}^{B^{\prime}}.
\end{equation}

The spinor $\psi_A$ represents the generator of the complementary
family to $\bar{\theta}^{B^{\prime}}$ of the sphere that intersect
$\Phi_{AB^{\prime}}$ in the same point as the
$\bar{\theta}^{B^{\prime}}-$generator, since
\begin{itemize}
    \item a linear relation must exist between these objects, due to the fact
    that generators of opposite type have a unique intersection.
    In fact, reconsidering the equation (\ref{psi}) we can rewrite it as
    \begin{equation}\label{zero psi}
        \Phi_{AB^{\prime}}\psi^A\bar{\theta}^{B^{\prime}}=0,
    \end{equation}
    having contracted both sides of equation (\ref{psi}) with $\psi^A$ and using $\psi_A\psi^A=0$ as in (\ref{prodotto interno zero}).
    Now, if $\Phi_{AB'}$ represents a plane, then
    $\psi^A\bar{\theta}^{B^{\prime}}$ represents a point
    $p^a$. This point is in general complex since the matrix obtained
    $P^{AB'}=\psi^A\bar{\theta}^{B^{\prime}}$ is only
    singular but not necessarily Hermitian as in (\ref{Xconspinore})
    or (\ref{Xcomplex}) and it is the intersection of the two generators (two lines) of the wave sphere of opposite families, $\psi^A$ and $\bar{\theta}^{B^{\prime}}$. Moreover, this point lies on the wave sphere since
    the matrix $P^{AB'}$ is singular. Finally, the relation (\ref{zero psi}) tells us
    that that this null point belongs to the plane
    $\Phi_{AB^{\prime}}$.
    \item (\ref{psi}) is the only such relation that transforms
    covariantly and homogeneously.
\end{itemize}
Now, the reason we claim that we can identify the wave sphere with
the Poincar\'{e} sphere is found by considering the effects of
pure rotations about the $k^a$ axis for any wave state: it is seen
that keeping $\bar{\theta}^{B'}$ fixed when the plane $\Phi_a$
rotates by $180^{\circ}$, the same point of intersection arises.
This explains the double rotation phenomenon of Stokes vectors,
when rotation takes place around the wave vector. In fact we can
construct the Stokes vector noting that only one point of any
generator is real. It is clear that
\begin{equation}\label{}
    \psi_A \bar{\psi}_{A^{\prime}}=\Psi_{AA'}= \begin{pmatrix}
     \:\psi_0\bar{\psi}_{0^{\prime}} & \psi_0\bar{\psi}_{1^{\prime}} \\
     \:\psi_1\bar{\psi}_{0^{\prime}} & \psi_1\bar{\psi}_{1^{\prime}} \\
    \end{pmatrix}\; \rightarrow \;S_a
\end{equation}
is a singular and Hermitian matrix (\ref{Xconspinore})
representing a coherency matrix for a single wave-state. It is
isomorphic to a real vector $S_a$, the Stokes vector\footnote{The
components of the Stokes vector appear to be different from the
usual ones, since we have used throughout this paper circular
polarization basis instead of the usual linear basis as stated in
the footnote \ref{nota basi}.}
\begin{eqnarray}\label{stokes circular basis}
    S&=&(S_0,S_1,S_2,S_3)= \\
    &=&(|\psi_0|^2+|\psi_1|^2,2\,\mbox{Re}[\psi_0\bar{\psi}_{1^{\prime}}],-2\,\mbox{Im}[\psi_0\bar{\psi}_{1^{\prime}}],|\psi_0|^2-|\psi_1|^2)
    \nonumber
\end{eqnarray}
whose discriminating condition
\begin{equation}\label{det di S}
    S_0^2-S_1^2-S_2^2-S_3^2=0
\end{equation}
recognizably represents a purely polarized state. Geometrically
the vanishing of the determinant of $\Psi_{AA'}$ expresses the
condition that a plane should contain two conjugate generators
($\psi_A$ and $\bar{\psi}_{A^{\prime}}$) of the polarization
sphere, so that the covariant Stokes vector $S_a$ represents a
real tangent plane of the Poincar\'{e} sphere which now is fused
with the wave sphere. In fact the condition (\ref{det di S}) can
be expressed as
\begin{equation}\label{}
    S_a\Omega^{ab}S_b=0
\end{equation}
which is the equation of a sphere in homogeneous coordinates (as
in (\ref{sfera onda}) and (\ref{sfera k})).

We finally remark that $S$ is not a true tensor object because,
like $\psi_A$ it omits structure (the phase flag essentially) that
also has to transform geometrically.

The introduction of the phase flag is the key element to identify
the polarization state for a wave vector in any direction with a
spinor. Any transformation of geometric orientation does not
change the spinor algebra. The spinor expression (\ref{psi})
transforms covariantly so that the relation will remain valid in
any rotated frame if all the elements composing the relation are
transformed with the appropriate rule as shown in (\ref{psi
tilde})-(\ref{teta}). The essential distinction between geometric
and polarimetric basis transformations is that in the former case
the phase flag $\bar{\theta}^{B^{\prime}}$ must be transformed for
consistency, while in the latter $\bar{\theta}^{B^{\prime}}$ must
be regarded as fixed. Fixing $\bar{\theta}^{B^{\prime}}$ and
varying $\Phi_{AB'}$ we obtain all the polarization states for one
direction of propagation.

\section{Illustrative examples}\label{HowToUseIt}
%How to use it

We will now report some numerical examples in order to illustrate
how all this really works.

We start first to choose one direction of propagation, which is as
usual the $z-$direction. The covariant $4-$vector $k_a$ is then
\begin{equation}\label{}
    k_a=k_0(1,0,0,-1)
\end{equation}
with $k_0=\frac{\omega}{c}$.
We choose now a linear polarized wave at $60^{\circ}$ to the $x-$axis propagating
along $z$. The electromagnetic field tensor will be
\begin{equation}\label{}
    F_{ab}=E\,\mbox{e}^{j(\omega t-\mathbf{k}\cdot \mathbf{x}+\alpha)}\begin{pmatrix}
  0 & \quad\!\cos\frac{\pi}{3} & \quad\!\sin\frac{\pi}{3} & 0 \\
  -\cos\frac{\pi}{3} & 0 & 0 & \cos\frac{\pi}{3} \\
  -\sin\frac{\pi}{3} & 0 & 0 & \sin\frac{\pi}{3} \\
  0 & -\cos\frac{\pi}{3} & -\sin\frac{\pi}{3} & 0 \\
\end{pmatrix}.
\end{equation}
where $E$ is the amplitude of the wave, $\omega t-\mathbf{k}\cdot
\mathbf{x}=\varphi$ is the phase (\ref{varphi scalare}) and
$\alpha$ is an initial phase.

In order to isolate $\Phi_a$ from $F_{ab}$, we use the (\ref{F})
contracting $F_{ab}$ with $t^b=(1,0,0,0)$. We obtain
\begin{equation}\label{phi con t}
  F_{ab}t^b=j(k_a\Phi_b-k_b\Phi_a)t^b \quad \Rightarrow \quad   F_{a0}=-j\,k_0\,\Phi_a
\end{equation}
and $\Phi_a$ has components
\begin{equation}\label{}
    \Phi_a=\frac{j}{k_0}\,E\,\mbox{e}^{j(\varphi+\alpha)}\,\left(0,-\cos\frac{\pi}{3},-\sin\frac{\pi}{3},0\right).
\end{equation}
Using (\ref{pot spinore}), the corresponding $\Phi_{AB'}$ is
\begin{equation}\label{}
    \Phi_{AB'}=\frac{j}{k_0}\,E\,\mbox{e}^{j(\varphi+\alpha)}\begin{pmatrix}
      0 & -\cos\frac{\pi}{3}+j\sin\frac{\pi}{3} \\
      -\cos\frac{\pi}{3}-j\sin\frac{\pi}{3} & 0 \\
    \end{pmatrix}.
\end{equation}
% NB Ho usato la convenzione di Klein: (1,e^{i\phi})->(1,cos\phi,\sin\phi,0)
Now after selecting a convenient fixed generator $\theta^{B'}$
\begin{equation}\label{thetaBprimo}
    \bar{\theta}^{B'}=\frac{1}{\sqrt{2}}\begin{pmatrix}
      1 \\
      1 \\
    \end{pmatrix},
\end{equation}
we are ready to compute the polarization spinor $\psi_A$ with the
(\ref{phitheta}):
\begin{eqnarray} \label{psi spinori}
     \psi_A&=&jk_0\,\Phi_{AB'}\bar{\theta}^{B'}=jk_0\,(\Phi_{A0'}\bar{\theta}^{0'}+\Phi_{A1'}\bar{\theta}^{1'})=
    \\
    &=&E\,\mbox{e}^{j(\varphi+\alpha)}\frac{1}{\sqrt{2}}\begin{pmatrix}
      \mbox{e}^{-j\frac{\pi}{3}} \\
      \mbox{e}^{+j\frac{\pi}{3}} \\
    \end{pmatrix}, \nonumber
\end{eqnarray}
which represents the Jones vector in circular polarization basis
for a $60^{\circ}$ linear polarization. The corresponding
polarization ratio is:
\begin{equation}\label{}
    \mu=\mbox{e}^{j\frac{2\pi}{3}}.
\end{equation}
 We can calculate the corresponding
Stokes vector using (\ref{stokes circular basis}) and we obtain:
\begin{equation}\label{}
    S_a=E^2
    \left(1,\cos\frac{2\pi}{3},\sin\frac{2\pi}{3},0\right).
\end{equation}
which is the Stokes vector for a linear polarization with
orientation angle $\psi=60^{\circ}$.

\begin{figure}[h]
\centering
\includegraphics[scale=0.4]{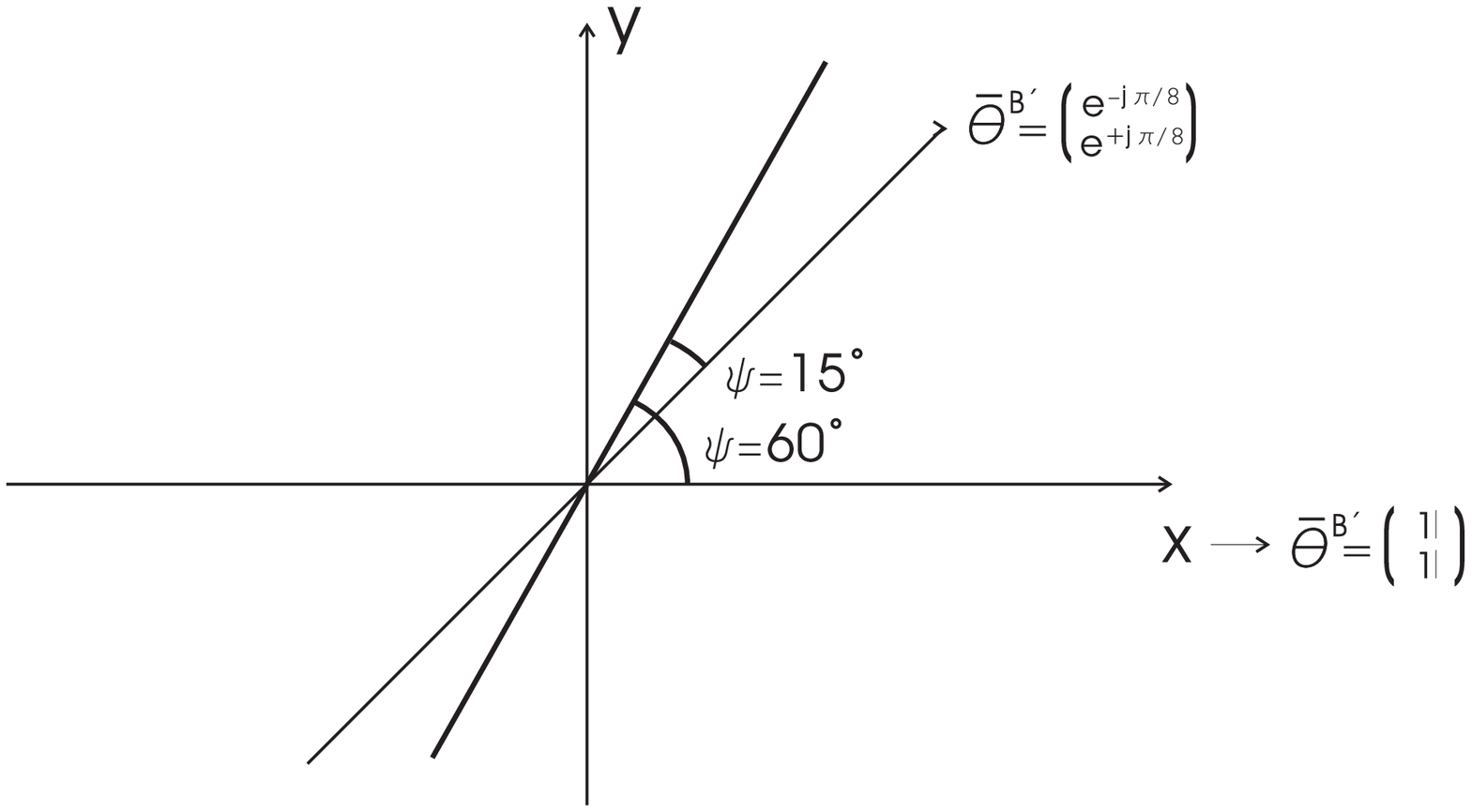}
 \caption{An example where the same linear polarization is defined using two different phase flags. In one case we have used $\bar{\theta}^{B'}=\textstyle{1 \choose 1}$ which is the $x-$axes in circular basis. The orientation angle is
$\psi=60^{\circ}$. In the other case we have used
$\bar{\theta}^{B'}=\textstyle{\;\mbox{e}^{-j\frac{\pi}{8}} \choose
\;\mbox{e}^{+j\frac{\pi}{8}}}$ which is an axis of $45^{\circ}$
respect to the $x-$axis. The resulting orientation angle is
$\psi=15^{\circ}$.} \label{linear60}
\end{figure}

The next step is to see what changes if we choose another
phase-flag $\bar{\theta}^{B'}$, keeping in mind (\ref{condizione su theta}). For
\begin{equation}\label{thetaBprimoI}
    \bar{\theta}^{B'}=\frac{1}{\sqrt{2}}\begin{pmatrix}
      \mbox{e}^{-j\frac{\pi}{8}} \\
      \mbox{e}^{j\frac{\pi}{8}} \\
    \end{pmatrix},
\end{equation}
we obtain
\begin{eqnarray} \label{psi spinoriI}
     \psi_A&=&jk_0\,\Phi_{AB'}\bar{\theta}^{B'}=jk_0\,(\Phi_{A0'}\bar{\theta}^{0'}+\Phi_{A1'}\bar{\theta}^{1'})=
    \\
    &=&E\,\mbox{e}^{j(\varphi+\alpha)}\frac{1}{\sqrt{2}}\begin{pmatrix}
      \mbox{e}^{-j(\frac{\pi}{3}-\frac{\pi}{8})} \\
      \mbox{e}^{+j\frac{\pi}{3}-\frac{\pi}{8}} \\
    \end{pmatrix}. \nonumber
\end{eqnarray}
The corresponding polarization ratio and Stokes vector are
\begin{equation}\label{}
    \mu=\mbox{e}^{2j(\frac{\pi}{3}-\frac{\pi}{8})}
\end{equation}
\begin{equation}\label{}
     S_a=E^2
    \left(1,\cos2(\frac{\pi}{3}-\frac{\pi}{8}),\sin2(\frac{\pi}{3}-\frac{\pi}{8}),0\right).
\end{equation}
which is a linear polarization with orientation angle
$\psi=15^{\circ}$. It is clear that the phase flag
$\bar{\theta}^{B'}$ keeps trace of the reference for the
orientation angle of the polarization ellipse (see Fig.
\ref{linear60}). We can notice that the components of the phase
flag $\bar{\theta}^{B'}$ apply phase factors to the circular
polarization components as we have shown in (\ref{psi nuova}). We
can conclude that this phase offset is relative to the origin of
the orientation angle. The phase flag is the element which defines
the orientation angle and the phases of the circular polarization
in a consistent manner. For these reasons it is the key to define
polarization states for arbitrarily oriented wave vectors.

Now we can do the same calculation using the tensors instead of
the spinors and for brevity we omit the multiplication
constants. We have the plane $\Phi_a$. We calculate the Pluecker
coordinates $\theta^{ab}$ of the line corresponding to the
generator $\bar{\theta}^{B'}$ and then the intersection of the
plane potential with the line generator, in order to find the
polarization state, the generator of the other type. We can use
(\ref{KAB con k}) where $\bar{\lambda}_{B'}=\displaystyle{1\choose
\lambda}$ and $\kappa_A=\displaystyle{1\choose \mu}$. We obtain:
\begin{equation}\label{generatori}
    \kappa_A\bar{\lambda}_{B'}=\begin{pmatrix}
  1 & \lambda \\
  \mu & \mu \lambda \\
\end{pmatrix}=\begin{pmatrix}
  \tau-z & -x+jy \\
  -x-jy & \tau+z \\
\end{pmatrix}.
\end{equation}
Since we want to compute the generator through $\bar{\theta}^{B'}$
then using (\ref{thetaBprimo}) $\lambda=-1$. The corresponding one
index tensor is:
\begin{equation}\label{}
    t^a=(1-\mu,1-\mu,j+j\mu,-1-\mu).
\end{equation}
In order to compute the projective Pluecker coordinates of this
line, we consider two points on this line for example
$p^a=(1,1,0,0)$ and $q^a=(0,0,j,-1)$,  and we use the relations
(\ref{skew line})-(\ref{plueckerCOORD}):
\begin{equation}\label{}
    \theta^{\,ab}=\begin{pmatrix}
  0 & 0 & j & -1 \\
  0 & 0 & j & -1 \\
  -j & -j & 0 & 0 \\
  1 & 1 & 0 & 0 \\
\end{pmatrix},
\end{equation}
\begin{equation}\label{}
    (0,-j,1,0,1,j).
\end{equation}
In order to find the generator of the other type, the polarization
state, we compute the intersection of the line $\theta^{\,ab}$
with the plane $\Phi_b$:
\begin{equation}\label{}
    \psi^a=\theta^{\,ab}\Phi_b=\left(\sin\frac{\pi}{3},\sin\frac{\pi}{3},-\cos\frac{\pi}{3},-j\cos\frac{\pi}{3}\right).
\end{equation}
The spinor corresponding to this point will be computed using
(\ref{generatori}):
\begin{equation}\label{}
    \mu=\frac{\tau+z}{-x+jy}=\mbox{e}^{2j\frac{\pi}{3}},
\end{equation}
with of course $\tau=\sin\frac{\pi}{3}$, $x=\sin\frac{\pi}{3}$,
$y=-\cos\frac{\pi}{3}$, $z=-j\cos\frac{\pi}{3}$. The spinor can be
expressed as:
\begin{equation}\label{}
    \psi_A\propto \begin{pmatrix}
      1 \\
      \mbox{e}^{j\frac{2\pi}{3}} \\
    \end{pmatrix}
\end{equation}
which is projectively the same as (\ref{psi spinori}).

Now we apply a rotation in space to the bivector $F^{ab}$, for
example a rotation of $45^{\circ}$ about the $y-$axis. The
rotation matrix turns out to be:
\begin{equation}\label{}
    R^{\,a}_{\;\; b}=\begin{pmatrix}
  1 & 0 & 0 & 0 \\
  0 & \cos\frac{\pi}{4} & 0 & -\sin\frac{\pi}{4} \\
  0 & 0 & 1 & 0 \\
  0 & \sin\frac{\pi}{4} & 0 & \cos\frac{\pi}{4} \\
\end{pmatrix}.
\end{equation}
The new electromagnetic field tensor will be:
\begin{equation}\label{F'con tensori}
    \tilde{F}^{cd}=R^{\,c}_{\;\; a}R^{\,d}_{\;\; b}\,F^{ab}=
\end{equation}
\begin{equation}
   \begin{pmatrix}
  0 & -\cos\frac{\pi}{4}\cos\frac{\pi}{3} & -\sin\frac{\pi}{3} & -\sin\frac{\pi}{4}\cos\frac{\pi}{3} \\
  \cos\frac{\pi}{4}\cos\frac{\pi}{3} & 0 & \sin\frac{\pi}{4}\sin\frac{\pi}{3} & \cos\frac{\pi}{3} \\
  \sin\frac{\pi}{3} & -\sin\frac{\pi}{4}\sin\frac{\pi}{3} & 0 & \cos\frac{\pi}{4}\sin\frac{\pi}{3} \\
  \sin\frac{\pi}{4}\cos\frac{\pi}{3} & -\cos\frac{\pi}{3} & -\cos\frac{\pi}{4}\sin\frac{\pi}{3} & 0 \\
\end{pmatrix}. \nonumber
\end{equation}
We can notice that the equivalent matrix calculation to
(\ref{F'con tensori}) would be:
\begin{equation}\label{F'matrice}
    \tilde{F}=RFR^{T}
\end{equation}
where $F$, $\tilde{F}$ and $R$ indicate the corresponding matrices
and $R^T$ is the transpose of the rotation matrix $R$.

The new wave vector will be:
\begin{equation}\label{}
    \tilde{k}^d=R^{\,d}_{\;\; a}k^a=\left(1,-\sin\frac{\pi}{4},0,\cos\frac{\pi}{4}\right)
\end{equation}
and
\begin{equation}\label{}
    \tilde{k}_a=\left(1,\sin\frac{\pi}{4},0,-\cos\frac{\pi}{4}\right).
\end{equation}
The new potential can be computed like in (\ref{phi con t}) and we
obtain:
\begin{equation}\label{}
    \tilde{\Phi}_a=\left(0,-j\cos\frac{\pi}{4}\cos\frac{\pi}{3},-j\sin\frac{\pi}{3},-j\sin\frac{\pi}{4}\cos\frac{\pi}{3}\right).
\end{equation}
We can easily verify that the potential is still a plane through
the origin $\omega^a$ and through the point $\tilde{k}^a$:
\begin{equation}\label{}
    \tilde{\Phi}_a\omega^a=0, \quad \tilde{\Phi}_a\tilde{k}^a=0
\end{equation}
In this calculation we are changing the orientation and
calculating the corresponding new polarization state. In order to
compute the new polarization state we have to rotate the generator
$\theta^{ab}$ as well, to obtain:
\begin{equation}\label{}
    \tilde{\theta}^{cd}=R^{\,c}_{\;\; a}R^{\,d}_{\;\;
    b}\,\theta^{ab}=\qquad \qquad \qquad \qquad \qquad\quad
\end{equation}
\begin{equation} \nonumber
    =\begin{pmatrix}
      0 & \sin\frac{\pi}{4} & j & -\cos\frac{\pi}{4} \\
      -\sin\frac{\pi}{4} & 0 & j\cos\frac{\pi}{4} & -1 \\
      -j & -j\cos\frac{\pi}{4} & 0 & -j\sin\frac{\pi}{4} \\
      \cos\frac{\pi}{4} & 1 & j\sin\frac{\pi}{4} & 0 \\
    \end{pmatrix}.
\end{equation}

In order to find the polarization state we need to find the point
where the line $\tilde{\theta}^{cd}$ intersects the plane
$\tilde{\Phi}_a$ as before\footnote{Of course
$\tilde{\psi}^a=R^{\,a}_{\;\; d}\psi^d$ but we have gone through
the full calculation again in order to make clear how tensors and
spinors work.}.
\begin{eqnarray}\label{}
    &&\tilde{\psi}^a=\tilde{\theta}^{ab}\tilde{\Phi}_b=\left(\sin\frac{\pi}{3},j\cos\frac{\pi}{3}\sin\frac{\pi}{4}+\sin\frac{\pi}{3}\cos\frac{\pi}{4},\right.
    \nonumber \\
    &&\left.-\cos\frac{\pi}{3},-j\cos\frac{\pi}{3}\cos\frac{\pi}{4}+\sin\frac{\pi}{3}\sin\frac{\pi}{4}\right).
\end{eqnarray}
Using the definition (\ref{generatori}) we can calculate the
polarization ratios of the two generators:
\begin{equation}\label{mug}
    \mu=\frac{\tau+z}{-x+jy}=\frac{\mbox{e}^{j\frac{2\pi}{3}}\cos\frac{\pi}{8}-\sin\frac{\pi}{8}}{\cos\frac{\pi}{8}+\mbox{e}^{j\frac{2\pi}{3}}\sin\frac{\pi}{8}},
\end{equation}
\begin{equation}\label{lambdag}
    \lambda=\frac{\tau+z}{-x-jy}=\frac{\cos\frac{\pi}{8}+\sin\frac{\pi}{8}}{-\cos\frac{\pi}{8}+\sin\frac{\pi}{8}}.
\end{equation}

Now we perform the same calculation using spinors. We want to calculate the new polarization state for a new orientation $\tilde{k}^a$. It will be much
faster and simpler. The unitary spinor corresponding to
$R^{\,a}_{\;\; b}$ is $U^{\,A}_{\;\; B}$:
\begin{eqnarray}\label{}
    U^{\,A}_{\;\; B}&=&-\cos\frac{\pi}{8}\,\epsilon^{\,A}_{\;\;
    B}-j\sin\frac{\pi}{8}(\tau^A \omega_B+\omega^A \tau_B)=  \nonumber\\
    &=&\begin{pmatrix}
      \cos\frac{\pi}{8} & \sin\frac{\pi}{8} \\
      -\sin\frac{\pi}{8} & \cos\frac{\pi}{8} \\
    \end{pmatrix}
\end{eqnarray}
where
\begin{equation}\label{}
    \tau^A=\frac{1}{\sqrt{2}}\begin{pmatrix}
  1 \\
  j \\
\end{pmatrix} \quad \omega^A=\frac{1}{\sqrt{2}}\begin{pmatrix}
  j \\
  1 \\
\end{pmatrix}
\end{equation}
are the spinor corresponding to the axis of rotation, the $y-$axis
$T_a=(1,0,1,0)$ and the orthogonal one $O_a=(1,0,-1,0)$. $U^{\,A}_{\;\; B}$ represents the rotation in space the generates the new orientation $\tilde{k}^a$.

The new polarization spinor can be simply computed as
\begin{equation}\label{}
   \tilde{\psi}^{A}=U^{\,A}_{\;\;
    B}\psi^{B}=\frac{j \,\mbox{e}^{-j\frac{\pi}{3}}}{\sqrt{2}}\begin{pmatrix}
  -\mbox{e}^{j\frac{2\pi}{3}}\cos\frac{\pi}{8}+\sin\frac{\pi}{8} \\
  \cos\frac{\pi}{8}+\mbox{e}^{j\frac{2\pi}{3}}\sin\frac{\pi}{8} \\
\end{pmatrix}.
\end{equation}
The corresponding covariant form of the spinor is:
\begin{equation}\label{}
    \tilde{\psi}_{A}=\frac{j \,\mbox{e}^{-j\frac{\pi}{3}}}{\sqrt{2}}\begin{pmatrix}
       -\cos\frac{\pi}{8}-\mbox{e}^{j\frac{2\pi}{3}}\sin\frac{\pi}{8} \\
  -\mbox{e}^{j\frac{2\pi}{3}}\cos\frac{\pi}{8}+\sin\frac{\pi}{8} \\
\end{pmatrix}
\end{equation}
and the corresponding polarization ratio is exactly (\ref{mug})
\begin{equation}\label{}
    \mu=\frac{\mbox{e}^{j\frac{2\pi}{3}}\cos\frac{\pi}{8}-\sin\frac{\pi}{8}}{\cos\frac{\pi}{8}+\mbox{e}^{j\frac{2\pi}{3}}\sin\frac{\pi}{8}}.
\end{equation}
The new phase flag $\bar{\tilde{\theta}}^{B'}$ can be simply
computed as
\begin{equation}\label{}
    \bar{\tilde{\theta}}^{A'}=\bar{U}^{\,A'}_{\;\;
    B'}\bar{\theta}^{B'}=\frac{1}{\sqrt{2}}\begin{pmatrix}
  \cos\frac{\pi}{8}+\sin\frac{\pi}{8} \\
  \cos\frac{\pi}{8}-\sin\frac{\pi}{8} \\
\end{pmatrix}.
\end{equation}
Using the conjugate version of (\ref{spinore covariante}), the
corresponding covariant phase flag is
\begin{equation}\label{}
    \bar{\tilde{\theta}}_{B'}=\epsilon_{A'B'}\bar{\tilde{\theta}}^{A'}=\frac{1}{\sqrt{2}}\begin{pmatrix}
  -\cos\frac{\pi}{8}+\sin\frac{\pi}{8} \\
   +\cos\frac{\pi}{8}+\sin\frac{\pi}{8}\\
\end{pmatrix}
\end{equation}
and the corresponding polarization ratio is (\ref{lambdag}).

The corresponding Stokes vectors are
\begin{equation}\label{}
    \tilde{\psi}^A \; \rightarrow \;
    \tilde{S}_a=\left(1,\cos\frac{2\pi}{3}\cos\frac{\pi}{4},\sin\frac{2\pi}{3},\cos\frac{2\pi}{3}\sin\frac{\pi}{4}\right),
\end{equation}
\begin{equation}\label{}
    \bar{\tilde{\theta}}^{B'}\; \rightarrow \;
    \Theta_b=\left(1,-\cos\frac{\pi}{4},0,-\sin\frac{\pi}{4}\right).
\end{equation}
$\tilde{\psi}^A$ is the new polarization state for a new orientation $\tilde{k}^a$ which corresponds to a new phase flag $\bar{\tilde{\theta}}^{B'}$.

\section{Discussion and conclusions}
%Discussion and conclusions

The representation that has been arrived at allows a spinor to
describe a polarization state for a wave vector in \emph{any}
direction. In any fixed direction, we extend to Jones vector
calculus in any chosen basis by applying unitary transformation to
the polarization spinor alone, and not to the phase flag. The
natural representation turns out, as might be expected, in terms
of a circular polarization basis. If we change the direction of
the wave vector we have also to apply the unitary transformation
to the phase flag.

Geometrically this representation can be determined since two
generators of the sphere $\psi_A$ and $\bar{\theta}^{B'}$ (one of
each kind) pass through any point of it, and then through every
tangential plane representing a wave vector there pass two
generators $\psi_A$ and $\bar{\psi}_{A'}$ which are complex
conjugates.

For the coherent field, one represents the polarization state of
the propagating wave as
\begin{equation}\label{}
    \psi_A\varpropto \mbox{e}^{jk_ax^a}=\mbox{e}^{j(\omega t-\mathbf{k}\cdot\mathbf{x})}
\end{equation}
and the other, the conjugate field as
\begin{equation}\label{}
    \bar{\psi}_{A^{\prime}}\varpropto \mbox{e}^{-jk_ax^a}=\mbox{e}^{j(-\omega
    t+\mathbf{k}\cdot\mathbf{x})}.
\end{equation}
These fields are conjugate solutions, but \emph{both} propagate in
the same direction, as the equations for constant phase surface
are identical. The use of strict spinor algebra prevents the
often-committed error of associating a conjugated Jones vector
with a backward propagating field. In each wave plane, therefore,
there is one generator for the unconjugated Fourier component, and
it is possible to show that the entire collection of these forms a
ruled surface (regulus) on the wave sphere. The generator lying in
the plane $(k_0,\mathbf{k})$ corresponds to the wave state of $-1$
helicity (LHC), while the generator in the plane corresponding to
the 'backward' direction $(k_0,-\mathbf{k})$ is that of $+1$
helicity (RHC). %MORE EXPLANATION!!!!!!

This is the origin of the apparent conjugation/reversal symmetry.
The lines on the regulus form a one-dimensional linear space, and
in this sense we can say
\begin{equation}\label{}
    \psi_A\equiv\begin{pmatrix}
      1 \\
      0 \\
    \end{pmatrix}+\psi\begin{pmatrix}
      0 \\
      1 \\
    \end{pmatrix} \quad \mbox{or} \quad \psi_A\equiv\psi_0\begin{pmatrix}
      1 \\
      0 \\
    \end{pmatrix}+\psi_1\begin{pmatrix}
      0 \\
      1 \\
    \end{pmatrix}
\end{equation}
for a homogeneous projective spinor in which the complex
polarization ratio $\psi=\frac{\psi_1}{\psi_0}$ runs from $0$ to
$\infty$. As we have shown there is an isomorphism between the
generators of the sphere and the set of the spinors $\psi_A$, and
this can be effected in an invariant way with respect to any
linear change of basis. Thus we have arrived at a unified
polarimetric description in which both the geometry of 'world
space' and the abstract mapping of the generators of the sphere
are handled consistently using spinors. The geometric
interpretation we have introduced explains the fundamental place
of the Poincar\'{e} sphere in polarimetry as an invariant object
under linear transformations with its reguli whose generators are
wave spinors, constituting its invariant subspaces. Identifying
generators as states of polarization, the structure of the
Poincar\'{e} sphere is preserved under all linear processes. The
sphere is considered an invariant of the theory, the 'absolute
quadric' \cite{semple}. The Poincar\'{e} sphere and the wave
sphere are hereby unified. The well-known phenomenon of 'double
rotation' of Stokes vectors with respect to rotation of world
coordinates is down to the fact that $\bar{\theta}^{B^{\prime}}$
is not rotated when a basis transformation is made while for
geometric transformations the phase flag is included (\ref{teta}).
It may appear that the phase flag, $\bar{\theta}^{B^{\prime}}$ was
introduced in a somewhat ad hoc manner, and indeed the choice
\begin{equation}\label{}
\bar{\theta}^{B^{\prime}}=\frac{1}{\sqrt{2}}\begin{pmatrix}
  1 \\
  1 \\
\end{pmatrix}
\end{equation}
was not a unique one. It is obvious that the absolute phase of
$\bar{\theta}^{B^{\prime}}$ could be chosen arbitrarily and even
that its elements undergo relative phase transformation without
loss of information. This is indeed the case, and so it appears
wholly reasonable that we name this object the \emph{phase flag}.
Its practical importance is that it defines a phase reference for
both components of the polarized wave. This is to be contrasted
with conventions fixing one component as real, eg
\begin{equation}\label{}
    \begin{pmatrix}
      \cos\theta \,\mbox{e}^{j\phi} \\
      \sin\theta \\
    \end{pmatrix},
\end{equation}
which fails when $\theta \; \rightarrow \; \frac{\pi}{2}$. In a
further planned paper on this topic we shall address the problems
of defining the phase of an electromagnetic plane wave, and
identifying the role of the phase flag in the relationship between
geometry and phase.

\section*{Acknowledgment}
% optional entry into table of contents (if used)
%\addcontentsline{toc}{section}{Acknowledgment}
This work was supported by the Marie Curie Research Training
Network (RTN) AMPER (Contract number HPRN-CT-2002-00205). The
authors are indebted to Prof. W-M. Boerner and Prof. M. Chandra
for their support and fruitful discussions. L.C. is also deeply
indebted to Prof. G. Wanielik for his inexaustible trust and
encouragement. The authors would also like to acknowledge the
diverse and invaluable discussions and debates with the late Dr.
Ernst L\"{u}neburg.

\bibliographystyle{IEEEtran}
\bibliography{IEEEabrv,bibliografia}

\vspace{-1cm}
\begin{biography}[{\includegraphics[width=1in,height=1.25in,clip,keepaspectratio]{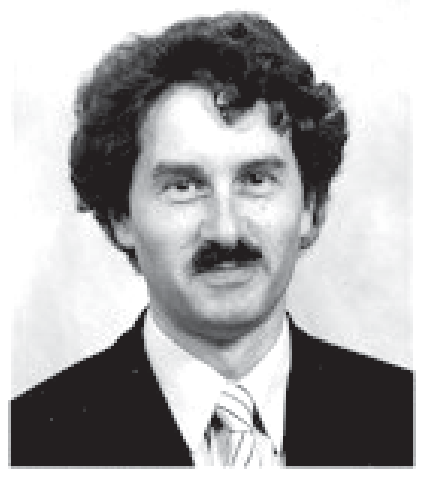}}]{David Bebbington}
Received the B.A. degree in experimental and theoretical physics
and the Ph.D. degree in radio astronomy from Cambridge University,
Cambridge, U.K., in 1977 and 1986, respectively. From 1981 to
1984, he worked on millimeter wave propagation research at the
Rutherford Appleton Laboratory. Since 1984, he has been at the
University of Essex, Essex, U.K., and currently holds the post of
Senior Lecturer in the Department of Computing and Electronic
System. His interests are polarimetry, applications of wave
propagation in remote sensing, and weather radars.
\end{biography}

\vspace{-1cm}

\begin{biography}[{\includegraphics[width=1in,height=1.25in,clip,keepaspectratio]{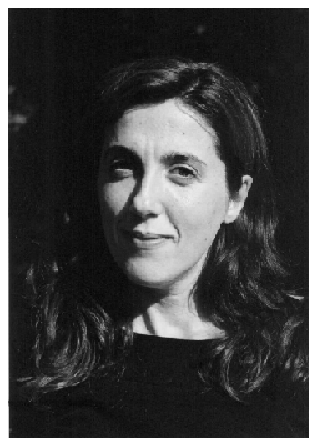}}]{Laura Carrea}
received the Laurea degree in physics from the University of
Turin, Italy in 1997. From 1999 to 2004 she worked at the Faculty
of Electrical Engineering and Information Technology at the
Chemnitz University of Technology in Germany partly with a
Fellowship in the frame of the Marie Curie TMR Network "Radar
Polarimetry: Theory and Application". Since 2004 she is with
Centre for Remote Sensing \& Environmetrics, Department of
Computing and Electronic System, University of Essex, U.K. Her
main interests center on radar polarimetry.
\end{biography}

\vspace{-0.8cm}

\begin{biography}[{\includegraphics[width=1in,height=1.25in,clip,keepaspectratio]{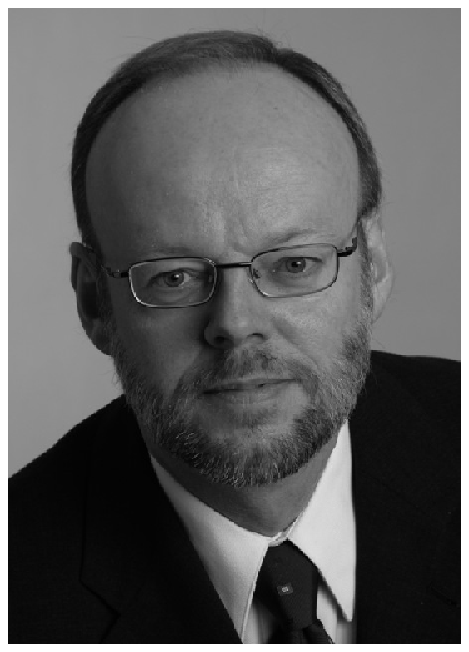}}]{Ernst Krogager}
received the M.Sc. degree in electrical engineering and the Danish
technical doctor (Dr.Techn.) degree from the Technical University
of Denmark, in 1981 and 1993, respectively. He has been an
employee of the Danish Defence Research Establishment since 1981.
Since 1984, he has conducted research in the areas of radar
non-cooperative target identification and polarimetric radar
imaging, and he has been the Danish representative in NATO working
groups on the subject. From November 1989 until June 1990, he was
temporarily with the Georgia Tech Research Institute, Atlanta,
Georgia, USA, for studying radar polarimetry. During the fall of
1993, he conducted postdoctoral studies in radar polarimetry at
the NASA Jet Propulsion Laboratory, Pasadena, California, USA. His
primary research interests currently include the subject of High
Power Microwaves.
\end{biography}

\newpage
\begin{table}[h]
%\centering
%\includegraphics[scale=0.4]{stereo}
\caption{} \label{table}
\end{table}
\vspace{-0.5cm}
\begin{tabular}{|c|c|c|}
  \hline
  & \multicolumn{2}{|c|}{\rule[-0.0cm]{0mm}{0.0cm}\textbf{The electromagnetic field}}  \\
  \hline
  \raisebox{6ex}{$\qquad F_{ab}=j(k_a\Phi_b-k_b\Phi_a) \qquad $} & \raisebox{6ex}{$F_{ab}=\begin{pmatrix}
  \quad\!0 & \quad\!E_x & \quad\!E_y & \quad\!0 \\
  -E_x & \quad\!0 & \!0 & \quad\!cB_y \\
  -E_y & \quad\!0 & \quad\!0 & -cB_x \\
  \!0 & -cB_y & \quad\!cB_x & \quad\!0 \\
\end{pmatrix}$} & \raisebox{2ex}{\includegraphics[scale=0.3]{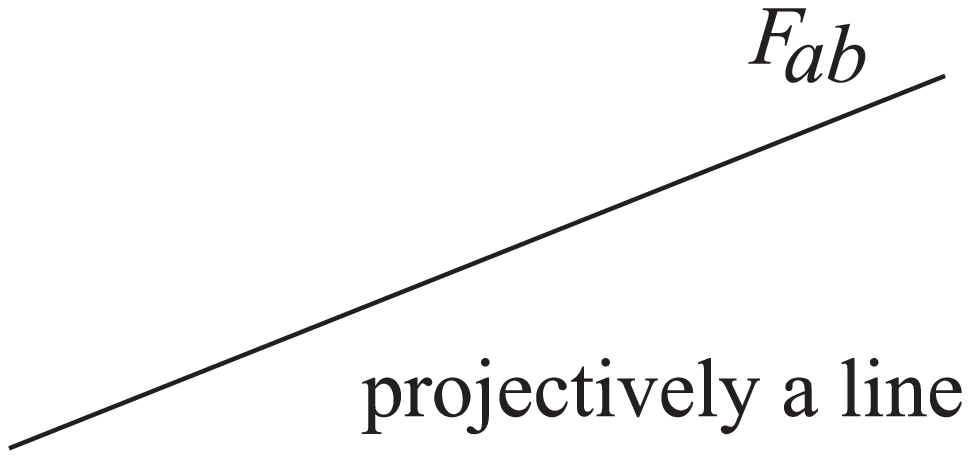}}  \\
  \cline{2-3}
  \multicolumn{3}{|l|}{\rule[-0.0cm]{0mm}{0.5cm}$\qquad$$\Bigg\downarrow \quad$ dropping
  the wave vector
  $k_a=(1,0,0,-1)$ information} \\
   \cline{2-3}
   & \multicolumn{2}{|c|}{\rule[-0.0cm]{0mm}{0.0cm}\textbf{The electromagnetic potential}}  \\
  \cline{2-3}
  \raisebox{11ex}{$\Phi_a=\frac{j}{k_0}F_{ab}t^b$} & \raisebox{11ex}{$\begin{array}{c}
    \Phi_a=\frac{j}{k_0}(0,-E_x,-E_y,0) \\
    \Phi_{AB'}=\frac{j}{k_0}\begin{pmatrix}
    0 & -E_x+jE_y \\
    -E_x-jE_y & 0 \\
  \end{pmatrix} \\
  \end{array}$} & \raisebox{0ex}{\includegraphics[scale=0.3]{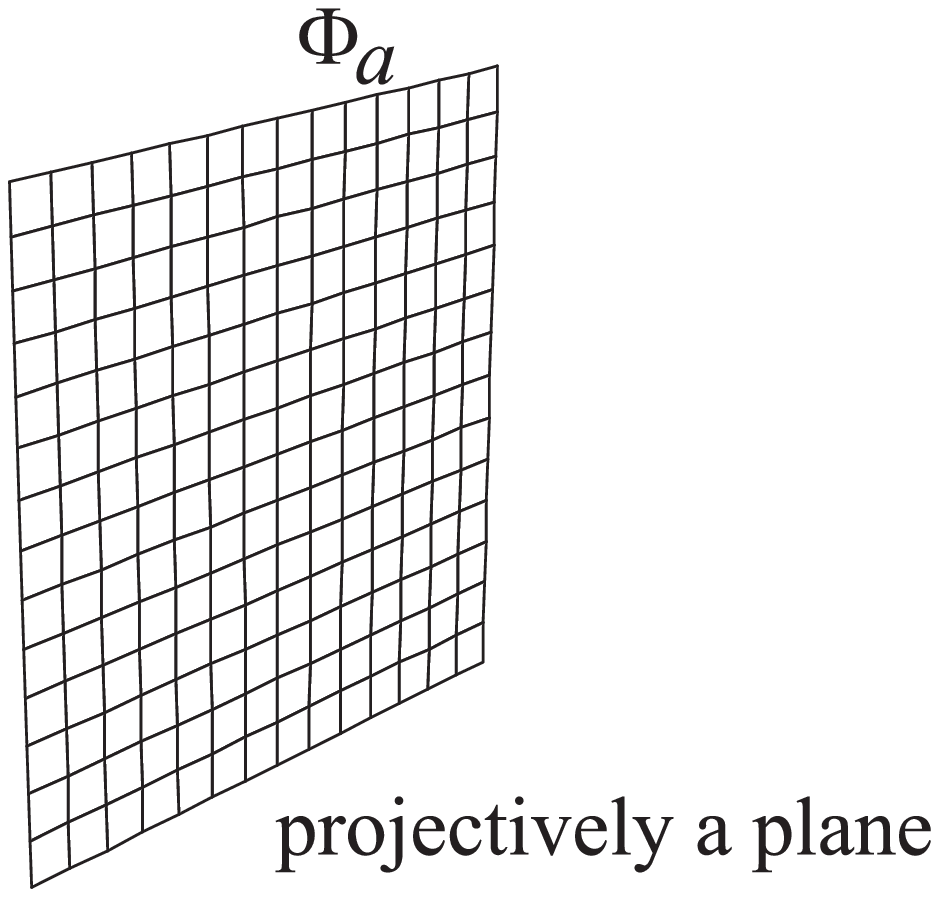}} \\
  \cline{2-3}
  \multicolumn{3}{|l|}{\rule[-0.0cm]{0mm}{0.5cm}$\qquad$$\Bigg\downarrow \quad$ dropping
  the reference
  $\bar{\theta}^{B'}=\displaystyle{1 \choose 1}$ information} \\
   \cline{2-3}
   & \multicolumn{2}{|c|}{\rule[-0.0cm]{0mm}{0.0cm}\textbf{The polarization spinor}} \\
  \cline{2-3}
  \raisebox{5ex}{$\psi_A=jk_0\Phi_{AB'}\bar{\theta}^{B'}$} &
\raisebox{5ex}{$\psi_A=\begin{pmatrix}
    E_x-jE_y \\
    E_x+jE_y \\
  \end{pmatrix}$} & \raisebox{-2ex}{\includegraphics[scale=0.3]{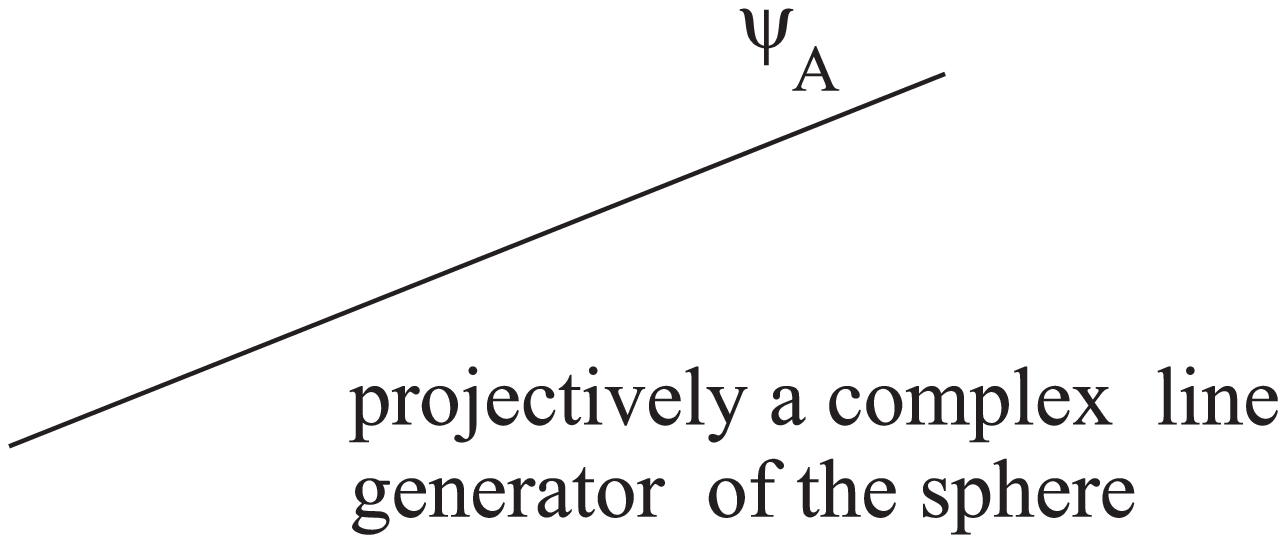}} \\
   \hline
  \end{tabular}

\newpage

\end{document}